\documentclass[lettersize,journal]{IEEEtran}
\usepackage{amsmath,amsfonts}
\usepackage{algorithmic}
\usepackage{array}
\usepackage[caption=false,font=normalsize,labelfont=sf,textfont=sf]{subfig}
\usepackage{textcomp}
\usepackage{stfloats}
\usepackage{url}
\usepackage{verbatim}
\usepackage{graphicx}

\usepackage[usenames,dvipsnames]{xcolor}
\usepackage{cite}
\newcommand{\amin}{A-Min$^*$}
\usepackage{bm}
\DeclareMathOperator*{\HD}{HD}
\DeclareMathOperator*{\sgn}{sgn}
\DeclareMathOperator*{\LUT}{LUT}

\usepackage{algorithmic}
\usepackage[linesnumbered,ruled,vlined]{algorithm2e}
\usepackage{amssymb,stmaryrd}
\usepackage{tikz}
\usepackage{boldline}
\usepackage{makecell}
\usepackage{threeparttable}
\usepackage{multirow}
\usepackage{booktabs}

\usepackage{pgfplots}
\usetikzlibrary{backgrounds}

\usepackage{calc}
\usepackage{colortbl}
\newcommand{\op}{
	\mathop{
		\vphantom{\bigoplus} 
		\mathchoice
		{\vcenter{\hbox{\resizebox{\widthof{$\displaystyle\bigoplus$}}{!}{$\boxplus$}}}}
		{\vcenter{\hbox{\resizebox{\widthof{$\bigoplus$}}{!}{$\boxplus$}}}}
		{\vcenter{\hbox{\resizebox{\widthof{$\scriptstyle\oplus$}}{!}{$\boxplus$}}}}
		{\vcenter{\hbox{\resizebox{\widthof{$\scriptscriptstyle\oplus$}}{!}{$\boxplus$}}}}
	}\displaylimits 
}
\newcommand{\bdc}{d_{c}}
\newcommand{\bdv}{d_{v}}
\definecolor{TableGray}{gray}{0.9}

\newcommand*\circled[1]{\tikz[baseline=(char.base)]{
		\node[shape=circle,draw,inner sep=0.25pt] (char) {#1};}}

\usepackage{balance}

\usepackage{siunitx}

\begin{document}
\title{A Generalized Adjusted Min-Sum Decoder for\\5G LDPC Codes: Algorithm and Implementation}
\author{Yuqing Ren,~\IEEEmembership{Student Member,~IEEE,}
	Hassan Harb,~\IEEEmembership{Member,~IEEE,}
	Yifei Shen,~\IEEEmembership{Member,~IEEE,}\\
	Alexios Balatsoukas-Stimming,~\IEEEmembership{Member,~IEEE,}
	and Andreas Burg,~\IEEEmembership{Senior Member,~IEEE}
	\thanks{Y. Ren, H. Harb, Y. Shen, and A. Burg are with the Telecommunications Circuits Laboratory (TCL), \'{E}cole Polytechnique F\'{e}d\'{e}rale de Lausanne (EPFL), Lausanne 1015, Switzerland (email: \{yuqing.ren, hassan.harb, yifei.shen, andreas.burg\}@epfl.ch). \emph{Corresponding author: Andreas Burg}.}
	\thanks{A. Balatsoukas-Stimming is with the Department of Electrical Engineering, Eindhoven University of Technology, 5600 MB Eindhoven, The Netherlands (email: a.k.balatsoukas.stimming@tue.nl).}
	\vspace{-0.25cm}
}


\maketitle

\begin{abstract}
	5G New Radio (NR) has stringent demands on both performance and complexity for the design of low-density parity-check (LDPC) decoding algorithms and corresponding VLSI implementations.
	Furthermore, decoders must fully support the wide range of all 5G~NR blocklengths and code rates, which is a significant challenge.
	In this paper, we present a high-performance and low-complexity LDPC decoder, tailor-made to fulfill the 5G requirements.
	First, to close the gap between belief propagation (BP) decoding and its approximations in hardware, we propose an extension of adjusted min-sum decoding, called generalized adjusted min-sum (GA-MS) decoding.
	This decoding algorithm flexibly truncates the incoming messages at the check node level and carefully approximates the non-linear functions of BP decoding to balance the error-rate and hardware complexity.
	Numerical results demonstrate that the proposed fixed-point GA-MS has only a minor gap of 0.1~dB compared to floating-point BP under various scenarios of 5G standard specifications.
	Secondly, we present a fully reconfigurable 5G~NR LDPC decoder implementation based on GA-MS decoding.
	Given that memory occupies a substantial portion of the decoder area, we adopt multiple data compression and approximation techniques to reduce 42.2\% of the memory overhead.
	The corresponding 28nm FD-SOI ASIC decoder has a core area of 1.823~mm$^{2}$ and operates at 895~MHz.
	It is compatible with all 5G~NR LDPC codes and achieves a peak throughput of 24.42~Gbps and a maximum area efficiency of 13.40~Gbps/mm$^{2}$ at 4 decoding iterations.
\end{abstract}


\section{Introduction}\label{sec:SecI_Intro}
\IEEEPARstart{L}{ow-density} parity-check (LDPC) codes, invented by Gallager~\cite{gallager1962low}, have received considerable attention in both academia and industry owing to their extraordinary error-correcting performance and the inherently parallel decoding algorithm.
Over the past several decades, LDPC codes have been adopted by various communication and storage systems, such as ATSC~\cite{ATSC2007}, IEEE~802.11n~\cite{80211n2008LDPC}, and {DVB-S2}~\cite{DVB2009}.
Most prominently, LDPC codes were ratified as the channel coding scheme of the enhanced mobile broadband (eMBB) scenario in 5G standards~\cite{5Gstandard2016, 5Gstandard2018}.
However, designing high-performance and low-complexity LDPC decoding algorithms and corresponding VLSI implementations~{tailored} to 5G New Radio (NR) is still an important research challenge.

In terms of decoding algorithms, belief propagation decoding (also called sum-product (SP) decoding on factor graphs~\cite{FR2001FGTIT}) of LDPC codes delivers outstanding error-correcting performance, closely approaching the Shannon limit~\cite{Urbanke2001TIT}.
However, SP decoding comes with high computational complexity and memory overhead~\cite{Mansour2003HTVLSI}.
To alleviate these issues, a series of min-sum (MS) decoding algorithms~\cite{wiberg1996codes, fossorier1999reduced, chen2005TCOM, wu2010adaptive, le2019adaptation, cui2020design, savin2008self, zhang2006two, kang2020enhanced} and a series of approximate-min$^*$ (\amin) decoding algorithms~\cite{jones2003approximate, zhou2019generalized} are proposed as alternatives to SP decoding.
For instance, MS decoding only involves selecting the smallest magnitudes among incoming messages from variable nodes (VNs) to check nodes (CNs), which simultaneously reduces the memory of outgoing messages from CNs to VNs.
However, this process results in performance loss compared to SP decoding~\cite{Adj2017OPatent, zhou2019generalized}.
Therefore, more advanced MS variants have been presented~\cite{chen2005TCOM, wu2010adaptive, le2019adaptation, cui2020design, savin2008self, zhang2006two, kang2020enhanced}, such as normalized MS (NMS) decoding, offset MS (OMS) decoding, adaptive MS (AMS) decoding~\cite{wu2010adaptive, le2019adaptation, cui2020design}, self-correction MS decoding~\cite{savin2008self}, and multiple dimensional modified MS decoding~\cite{zhang2006two, kang2020enhanced}.

On the other hand, \amin~decoding first determines the edge with the smallest incoming message, then calculates two distinct outgoing magnitudes at each CN, and propagates them to adjacent VNs,  relying on identical SP functions~\cite{jones2003approximate}.
While this method nearly matches the performance of SP decoding, \amin\;decoding suffers from two significant drawbacks: long decoding latency from the above sequential processing and substantial computational complexity due to the SP functions.
To mitigate the decoding latency issue, the authors of~\cite{zhou2019generalized} proposed a generalized \amin\;(G\amin) decoding algorithm by truncating the number of incoming messages to optimize recursive CN processing.
To reduce computational complexity, adjusted MS (A-MS) decoding proposed by Qualcomm~\cite{Adj2017OPatent} has drawn significant attention during the development of 5G~NR LDPC codes.
A-MS decoding can be considered as a quantized version of \amin\;decoding, as it employs look-up tables (LUTs) to simplify non-linear CN processing.
However, compared to classical MS-based decoders~\cite{kuo2008flexible, zhang2010JSSC, Studer2008Asilomar, Roth2010ASSCC}, A-MS decoding still faces a relatively high implementation complexity due to extra additions and comparisons in the approximation process.

In terms of hardware implementations and given that the 5G standard stipulates a peak throughput of $20$~Gbps in the downlink~\cite{Hui2018VTMag}, the 5G~NR LDPC decoder is tasked with balancing throughput, area efficiency, and energy consumption.
This balance must also uphold compatibility across all code configurations, presenting a significant challenge.
Notably, since the large amount of memory required to support the maximum blocklength in 5G~NR already occupies a significant part of the decoder area, these large memories tend to reduce the impact of more complex algorithms on overall efficiency.
In the literature~\cite{blanksby2002690, Mansour2003HTVLSI, cheng2014fully, ghanaatian2017588, kuo2008flexible, zhang2010JSSC, Studer2008Asilomar, Roth2010ASSCC}, numerous classical LDPC decoders have been presented, featuring varying degrees of resource sharing.
These decoders can be categorized into fully-parallel~\cite{blanksby2002690, Mansour2003HTVLSI, cheng2014fully, ghanaatian2017588} and partially-parallel architectures~\cite{kuo2008flexible, zhang2010JSSC, Studer2008Asilomar, Roth2010ASSCC}.
Partially-parallel decoders can be further divided into block-parallel, row-parallel, and other variants.
It is noteworthy that thanks to the quasi-cyclic (QC) property of the LDPC codes in many standards~\cite{80211n2008LDPC,5Gstandard2016,5Gstandard2018}, block-parallel architectures can implement LDPC decoding in an iteratively decomposed fashion without high routing complexity, resulting in a balance between throughput, area efficiency, and decoding flexibility~\cite{Studer2008Asilomar, Roth2010ASSCC}.

To adhere to the 5G peak throughput requirement, several state-of-the-art (SOA) 5G~NR LDPC decoders have been reported in~\cite{cui2020design, lin202133Gbps, Lee2022TCASI, nadal2021parallel, yun2021area, Su2022SSCL, verma2022low}, mainly using row-parallel architectures or variations thereof.
Unlike block-parallel ones, these row-parallel architectures can process multiple blocks simultaneously by a more complex programmable routing network to improve peak throughput.
However, due to the high routing complexity for long codes, it is difficult for these designs to support the maximum blocklength in 5G~NR.
Thus, the implementations in~\cite{cui2020design, nadal2021parallel, yun2021area, Su2022SSCL, verma2022low} generally target or provide results only for short to moderate blocklengths.
Only recently such a 5G~NR LDPC decoder with a novel memory access scheduling~\cite{Lee2022TCASI} was reported to fully meet the requirements of 5G.
Based on partially row-parallel (PRP) architecture, the decoder in~\cite{Lee2022TCASI} can process each layer of the 5G~NR LDPC base graphs within a predetermined fixed latency to achieve a high peak throughput.
Nevertheless, the row weights of the base graphs vary significantly, and most layers have low weights~\cite{5Gstandard2016,5Gstandard2018}.
During decoding at medium to low code rates, the efficiency of this PRP architecture~\cite{Lee2022TCASI} is limited by its row-parallel design that must still process layers with low weights sequentially at a fixed rate, thus diminishing the decoding throughput.
Yet, considering the wide range of 5G~NR LDPC blocklengths and code rates, the block-parallel architecture has an inherent advantage in balancing flexibility and parallelism, achieving stable performance across~all 5G~NR code configurations.
Our work demonstrates that a block-parallel architecture can already satisfy the 5G peak throughput requirement of $20$~Gbps, without higher processing~parallelism.

\subsection*{Contributions and Paper Outline:}
The specific contributions of this paper are as follows:

\begin{itemize}
	\item We propose a high-performance and low-complexity algorithm called generalized A-MS (GA-MS) decoding to balance the error-rate, computational complexity, and~memory overhead.
	We also design the required LUTs and propose other optimizations to simplify the hardware.
	
	\item We provide a comprehensive performance analysis with the designed LUTs, quantization schemes, and approximation techniques. 
	The proposed fixed-point GA-MS decoding has only a $0.1$~dB gap compared to floating-point SP decoding under various scenarios of 5G~NR standard specifications.
	
	\item A hardware-friendly optimized static schedule (OSS) is proposed to both improve error-correcting performance and reduce the worst-case decoding latency.
	
	\item We present a fully reconfigurable 5G~NR LDPC decoder implementation using GA-MS decoding, compatible with all 5G~NR LDPC codes.
	By adding data compression and approximation techniques, we achieve a significant reduction in memory overhead compared to explicit storage.
	The $28$nm FD-SOI post-layout implementation has a core area of $1.823$~mm$^{2}$, achieves a peak throughput of $24.42$~Gbps at $895$~MHz, and has an energy consumption of $12.56$~pJ/bit with a supply voltage of $1.0$~V.
\end{itemize}
The remainder of this paper is organized as follows: Section~\ref{sec:Preli} provides symbol definitions and background on LDPC codes and decoding.
Section~\ref{sec:SecIII_GAMS} describes the proposed GA-MS decoding and its various optimizations.
In Section~\ref{sec:SecIV_all}, we present our 5G~NR LDPC decoder architecture and the corresponding optimizations.
Section~\ref{sec:SecV_synthesis} discusses implementation results.
Finally, Section~\ref{sec:SecVI_conclusion} concludes the paper.

\section{Preliminaries}\label{sec:Preli}
\begin{figure}[t]
	\centering
	\includegraphics[width=0.8\linewidth]{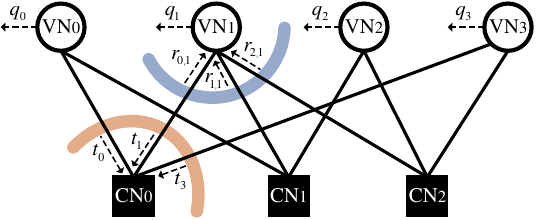}
	\caption{A bipartite Tanner graph with VN update and CN update process.}
	\label{fig:Tanner}
\end{figure}

Throughout this paper, we follow the definitions introduced below.
Boldface small letters such as $\bm{u}$ denote vectors, where $\bm{u}[i]$ refers to the $i$-th element of $\bm{u}$.
Boldface capital letters such as $\mathbf{B}$ represent matrices, where $\mathbf{B}[i][j]$ denotes the element at the $i$-th row of the $j$-th column of~$\mathbf{B}$.
Blackboard letters such as $\mathbb{S}=\{\cdot\}$ denote sets with $|\mathbb{S}|$ being the cardinality of $\mathbb{S}$.
The hard decision function is defined as $\HD(x)=1$ if $x<0$ and $\HD(x)=0$ if $x\geq 0$.
The signum function, denoted as $\sgn(x)$, returns $-1$, $0$, or $1$, when $x$ is negative, zero, or positive, respectively.
For brevity, we use the term \emph{floating-point} to specifically refer to double-precision floating-point.
\begin{equation}\label{eq:ldpc_def}
	\left\{\bm{x}\in\{0,1\}^{N}|\mathbf{H}\cdot\bm{x}=\bm{0}_{M\times 1}\right\},
\end{equation}
LDPC codes are linear block codes specified by a sparse $M\times N$ parity-check matrix (PCM) $\mathbf{H}$, as~shown~in~\eqref{eq:ldpc_def}, where $\bm{x}$ is a length-$N$ binary column vector and $\bm{0}$ is a length-$M$ all-zero~vector.
$M$ denotes the number of parity checks and $N$ denotes the code length.
Furthermore, LDPC codes can also be described by a bipartite Tanner graph with a set of $M$ CNs and $N$ VNs~\cite{tanner1981recursive}.
If $\mathbf{H}[c][v]=1$ for $0\leq c<M$ and $0\leq v<N$, the $c$-th CN is connected to the $v$-th VN on the Tanner graph.
For each CN, we use $\mathbb{V}_c$ to represent the set of the adjacent VNs for the $c$-th CN and we use $\mathbb{C}_{v}$ to denote the set of the neighbours of the $v$-th VN.
The number of neighbours that connect to a VN or a CN are referred to as their column- and row-degree, denoted as $d_v$ and $d_c$, i.e., $|\mathbb{C}_v|=d_v$ and $|\mathbb{V}_{c}|=d_c$.
Due to the quasi-cyclic property~\cite{fossorier2004quasicyclic, zhong2005block}, QC-LDPC codes are further described by a more structured $M_p\times N_p$ prototype matrix $\mathbf{H}_{p}$.
Each entry of $\mathbf{H}_p$ can be expanded by substituting each element of $\mathbf{H}_p$ with a $Z\times Z$ identity matrix that is cyclically shifted by $\omega=\mathbf{H}_{p}[c][v]<Z$ for $0\leq c<M_p$ and $0\leq v< N_p$.
The code parameter $Z$ is referred to as the \emph{lifting size}.
If $\omega=-1$ or $\omega=0$, the corresponding entry denotes a $Z\times Z$ all-zero~or~identity~matrix,~respectively.

\subsection{5G~NR LDPC Codes}\label{sec:SecII_5GLDPC}
To enable rate-compatibility and incremental redundancy hybrid automatic repeat request (IR-HARQ), 5G~NR adopts protograph-based raptor-like LDPC codes.
Combined with the quasi-cyclic property, 5G~NR LDPC codes can be derived from two base graphs (BG1 and BG2).
Let $K_u$ denote the number of information columns in the base graphs.
In 5G standards, a complete BG1 comprises $46$ rows and $68$ columns (with the maximum $K_u=22$), and a complete BG2 consists of $42$ rows and $52$ columns (with the maximum $K_u=10$).
A variety of code lengths and rates are attained by adjusting the lifting size $Z$ and by puncturing the columns of the base graphs~\cite{5Gstandard2018,Zhon2020TCASII},\footnote{In 5G standards, the lifting size $Z$ covers $51$ distinct values ranging from $2$ to $384$ ($Z=i\times 2^{n}$, $i\in\{2,3,5,7,9,11,13,15\}$, $n$ are non-negative integers). For further details, please refer to Table 5.3.2-1 in~\cite{5Gstandard2018}.} which affects the number of parity-check bits.
The leftmost two information columns in~the~base~graphs are always punctured to boost transmission efficiency in practical~scenarios.
Let $E$ and $K$ denote the actual transmitted code length and information length after rate-matching.
We can refer to 5G~NR LDPC codes as $(E,K)$ codes, where $K=K_{u}\cdot Z$ and where $R=\frac{K}{E}$ represents~the~code~rate.

\subsection{Layered LDPC Decoding}\label{ref:Sec2_layered}
The error-correcting performance and hardware complexity of an LDPC decoder also depend on its decoding schedule.
Two classical methods are flooding~\cite{Urbanke2001TIT} and layered schedules~\cite{sharon2004efficient, hocevar2004reduced}.
In contrast to the flooding schedule that updates all VNs together at the end of each iteration~\cite{Urbanke2001TIT}, layered decoding goes through the PCM and updates the connected VNs row by row.
This update strategy results in faster convergence and a significant reduction in memory~overhead, as fewer messages from VNs to CNs need to be stored.

Let $q_{v}$ denote the \emph{posterior} log-likelihood ratio (LLR) associated with the $v$-th VN, which is the aggregate value of all incoming messages and the channel LLR $y_{v}$. 
Let $r_{c,v}$ denote the message from the $c$-th CN to the $v$-th VN.
We also define an intermediate variable $t_{v}$, corresponding to the message from the $v$-th VN at the current row.
Layered decoding is completely defined by the aforementioned three message types: Q-message $q_{v}$, R-message $r_{c,v}$, and T-message $t_{v}$, as shown in Fig.~\ref{fig:Tanner}.
When processing the $c$-th row (i.e., the $c$-th CN) at the $i$-th iteration, layered decoding executes the updates shown in~\eqref{eq:layeredMS} (taking MS decoding as an example).
Each of the three equations in~\eqref{eq:layeredMS} is only performed after the previous equation has been evaluated for all VNs, $0\leq v<N$. 
\begin{subequations}\label{eq:layeredMS}
	\begin{align}
		&t_{v}\;\;\leftarrow\;q_{v}-r_{c,v},\label{eq:21}\\
		&r_{c,v}\leftarrow\prod_{v'\in \mathbb{V}_{c}\backslash v}\sgn(t_{v'})\cdot\min\limits_{v'\in \mathbb{V}_{c}\backslash v}(|t_{v'}|),\label{eq:22}\\
		&q_{v}\;\;\leftarrow\;t_{v}+r_{c,v}.\label{eq:23}
	\end{align}
\end{subequations}

First, the intermediate $t_{v}$ values are computed on the fly using the stored $q_{v}$ and $r_{c,v}$.
Subsequently, the minimum and sign are selected, excluding the message along the current edge itself, to update all $q_{v}$ and $r_{c,v}$ values.
At the beginning of the first iteration, the $q_{v}$ for $0\leq v<N$ are initialized by the channel LLRs $y_{v}$ and the $r_{c,v}$ for $0\leq c<M$ and $0\leq v<N$ are set to zero. 
Once the maximum decoding iterations $I_{\max}$ are reached, the tentative codeword can be obtained by $\hat{x}_{v}=\HD(q_{v}),0\leq v<N$.
For 5G~NR LDPC codes, the $c$-th row of $\mathbf{H}_{p}$ (i.e., BG1 and BG2) corresponds to the rows $c\cdot Z $ to $(c+1)\cdot Z-1$ of $\mathbf{H}$.
For simplicity, we call the rows $c\cdot Z $ to $(c+1)\cdot Z-1$ of $\mathbf{H}$ the $c$-th~layer.

\subsection{Adjusted Min-Sum (A-MS) Decoding}
In \amin~decoding, the smallest magnitude of all incoming T-messages is identified first and two distinct outgoing magnitudes are computed using~\eqref{eq:A-MSbasic}.
Unlike MS decoding in~\eqref{eq:layeredMS}, \amin~decoding still requires complex SP calculations on all incoming T-messages after finding the minimum, which can result in a long decoding latency.
However, based on the block-parallel architecture, using the same equations~\eqref{eq:A-MSbasic}, A-MS decoding~\cite{Adj2017OPatent} can select the second smallest value when a new $t_{v}$ arrives, and perform a box-plus operator with the new second minimum and the previous result to recursively update the outcome.
This approach has the advantage that both results of~\eqref{eq:A-MSbasic} and the minimum can be obtained simultaneously after the arrival of the last valid block in the current row to avoid serial processing.
The box-plus operator is approximated using simple LUTs to reduce hardware~complexity.
Corresponding to the edge with the minimum incoming magnitude, the outgoing R-message is referred to as a \emph{critical message}~\cite{zhou2019generalized} (consistent with the principle of SP decoding).
For the remaining edges, the outgoing R-messages are called \emph{non-critical messages} and are computed using all incoming messages (including the current edge itself).
Hence, as a hardware-friendly decoding algorithm, A-MS decoding has a similar storage complexity as MS-based decoding and can achieve almost the same error-correcting~performance~as~SP~decoding.
\begin{equation}\label{eq:A-MSbasic}
	r_{c,v}\leftarrow\left\{
	\begin{aligned}
		&\prod_{v'\in \mathbb{V}_{c}\backslash v}\!\!\sgn(t_{v'})\cdot\!\!\!\!\op_{v'\in \mathbb{V}_{c}\backslash v}\!\!|t_{v'}|,\;\mathrm{if\;}t_{v}\mathrm{\;is\;minimum},\\
		&\prod_{v'\in \mathbb{V}_{c}\backslash v}\!\!\sgn(t_{v'})\cdot\!\op_{v\in \mathbb{V}_{c}}|t_{v}|,\;\mathrm{otherwise}.
	\end{aligned}
	\right.
\end{equation}

Despite its superiority, conventional A-MS decoding still suffers from several points to be optimized.
First, the influence of most incoming T-messages on the final outgoing R-messages is negligible based on our simulations, which means that most of the box-plus operators in~\eqref{eq:A-MSbasic} can be skipped to save computational complexity.
In A-MS decoding, the identification of the minimum and the computation of the R-messages are tightly linked.
This process relies on the block-parallel architecture that processes the incoming message block-by-block to resolve the data dependency, limiting its potential for increased parallelism.
In addition, a straightforward way to design the box-plus operator in A-MS decoding is to approximate it as multiple serial small LUTs (as referred to~\cite{Adj2017OPatent}), leading to a relatively long data path due to the comparison, addition, and~LUT~operations.

\section{Proposed High-Performance and Low-Complexity Decoding Algorithms}\label{sec:SecIII_GAMS}
In this section, we propose a novel algorithm called GA-MS decoding, by extending the above A-MS decoding algorithm~to a generalized form.
Our algorithm offers high-performance and low-complexity decoding.
By truncating the number of incoming messages in the CN processing, we can make a trade-off between the error-rate and computational complexity.
Moreover, combined with well-designed LUTs, quantization schemes, and approximation techniques, we provide a comprehensive performance analysis of fixed-point GA-MS decoding to demonstrate its stable and good error-rate across various code configurations and high-order~modulations.

\subsection{Generalized Adjusted Min-Sum (GA-MS) Decoding}\label{sec:SecIII_A}
Let $t_{a}$ and $t_{b}$ denote two arbitrary incoming T-messages.
The complete box-plus operator between $t_{a}$ and $t_{b}$ is:
\begin{equation}\label{eq:boxplus}
	\begin{aligned}
		t_{a}\op t_{b}&=2\tanh^{-1}\!\left(\tanh\left(\frac{t_{a}}{2}\right)\cdot\tanh\left(\frac{t_{b}}{2}\right)\right)\!\\
		&=\!\sgn(t_{a})\!\cdot\!\sgn(t_{b})\!\cdot\!\left(\!\!
		\begin{aligned}
			&\min\left(|t_{a}|, |t_{b}|\right)\\
			+&\ln\left(1+e^{-\left||t_{a}|+|t_{b}|\right|}\right)\\
			-&\ln\left(1+e^{-\left||t_{a}|-|t_{b}|\right|}\right)\\
		\end{aligned}
		\!\!\right)\\
		&\leq\sgn(t_{a})\cdot\sgn(t_{b})\cdot\min\left(|t_{a}|, |t_{b}|\right),\\
	\end{aligned}
\end{equation}
where the negative non-linear term $\triangle(t_{a}, t_{b}):=\ln\left(1+e^{-\left||t_{a}|+|t_{b}|\right|}\right)-\ln\left(1+e^{-\left||t_{a}|-|t_{b}|\right|}\right)$ is the reason why MS decoding always overestimates SP decoding~\cite{wiberg1996codes, fossorier1999reduced}.
If $|t_{a}|\ll|t_{b}|$, we can further simplify~\eqref{eq:boxplus} as
\begin{equation}\label{eq:boxplusMS}
	t_{a}\op t_{b}\approx\sgn(t_{a})\cdot\sgn(t_{b})\cdot|t_{a}|,
\end{equation}
which means that the smallest incoming magnitude dominates in~\eqref{eq:boxplus}.
Hence, in CN processing, more emphasis is placed on these incoming T-messages with smaller magnitudes, assigning them greater weights in the outgoing numerical calculation.
Similar to~\cite{zhou2019generalized}, we define a new set called $\mathbb{V}^{\gamma}_{c}$ to only contain the first $\gamma$ smallest magnitude values.
Note that if $d_c\geq\gamma$, $|\mathbb{V}^{\gamma}_{c}|=\gamma$ holds, but if $d_c<\gamma$, $\mathbb{V}^{\gamma}_{c}$ is equivalent to $\mathbb{V}_{c}$ without any message truncation (i.e., $|\mathbb{V}^{\gamma}_{c}|=|\mathbb{V}_{c}|=d_c$).
The corresponding update of~\eqref{eq:A-MSbasic} is as follows, where we can flexibly configure the parameter $\gamma$ to adjust the number of incoming messages used in CN processing.
\begin{equation}\label{eq:GA-MSbasic}
	r_{c,v}\leftarrow\left\{
	\begin{aligned}
		&\!\prod_{v'\in \mathbb{V}_{c}\backslash v}\!\sgn(t_{v'})\cdot\!\!\!\!\op_{\tilde{v}^{'}\in \mathbb{V}^{\gamma}_{c}\backslash v}\!\!\!|t_{\tilde{v}^{'}}|,\;\mathrm{if\;}t_{v}\mathrm{\;is\;minimum},\\
		&\!\prod_{v'\in \mathbb{V}_{c}\backslash v}\!\sgn(t_{v'})\cdot\op_{\tilde{v}\in \mathbb{V}^{\gamma}_{c}}|t_{\tilde{v}}|,\;\mathrm{otherwise}.
	\end{aligned}
	\right.
\end{equation}

As shown in~\eqref{eq:GA-MSPart}, $|t_{v^{*}}|$ is the $(\gamma+1)$-th smallest~incoming~magnitude.
It is intuitive to prove that when $\gamma$ is larger, the outgoing $r_{c,v}$ approaches the original A-Min* result. 
\begin{equation}\label{eq:GA-MSPart}
	\begin{aligned}
		\op\limits_{\tilde{v}\in \mathbb{V}^{\gamma}_{c}}|t_{\tilde{v}}|\geq\!\left(\op\limits_{\tilde{v}\in\mathbb{V}^{\gamma}_{c}}|t_{\tilde{v}}|\right)\op|t_{v^{*}}|=\!\!\!\!\op\limits_{\tilde{v}\in\mathbb{V}^{\gamma+1}_{c}}\!\!|t_{\tilde{v}}|\geq\hdots\geq\!\op\limits_{v\in \mathbb{V}_{c}}\!|t_{v}|.
	\end{aligned}
\end{equation}

However, if the value of $\gamma$ is relatively small, GA-MS decoding truncates too much information, resulting in an obvious performance loss.
In order to compensate for this degradation, we introduce an additional offset $\beta$ in~\eqref{eq:GA-MSadvanced} to reasonably scale the outgoing R-message and effectively alleviate the~overestimation~phenomenon as
\begin{equation}\label{eq:GA-MSadvanced}
	r_{c,v}\leftarrow\!\left\{
	\begin{aligned}
		&\!\!\prod_{v'\in \mathbb{V}_{c}\backslash v}\!\!\sgn(t_{v'})\cdot\!\max\!\left(\!\left(\op_{\tilde{v}^{'}\in \mathbb{V}^{\gamma}_{c}\backslash v}\!\!|t_{\tilde{v}^{'}}|\right)\!-\!\beta,0\right)\!,\\
		&\mathrm{if\;}t_{v}\mathrm{\;is\;minimum},\\
		&\!\!\prod_{v'\in \mathbb{V}_{c}\backslash v}\!\!\sgn(t_{v'})\cdot\!\max\!\left(\!\left(\op_{\tilde{v}\in \mathbb{V}^{\gamma}_{c}}|t_{\tilde{v}}|\right)-\beta,0\right)\!,\\
		&\mathrm{otherwise}.\\
	\end{aligned}
	\right.
\end{equation}

It is worth noting that, compared to the original A-MS decoding algorithm~\cite{Adj2017OPatent}, the proposed GA-MS decoding has a completely different decoding process, which first collects the $\gamma$ smallest incoming magnitudes just as the MS-based decoders do and then executes $\gamma-1$ approximate box-plus operators and an additional subtraction together in the CN processing.
The above improvement endows GA-MS decoding with a similar efficient hardware architecture as classical MS-based decoders~\cite{Studer2008Asilomar, Roth2010ASSCC}, which significantly facilitates the corresponding decoder~implementation (discussed in Section~\ref{sec:SecIV_all}).

\begin{figure}[t]
	\centering
	\includegraphics[width=0.95\columnwidth]{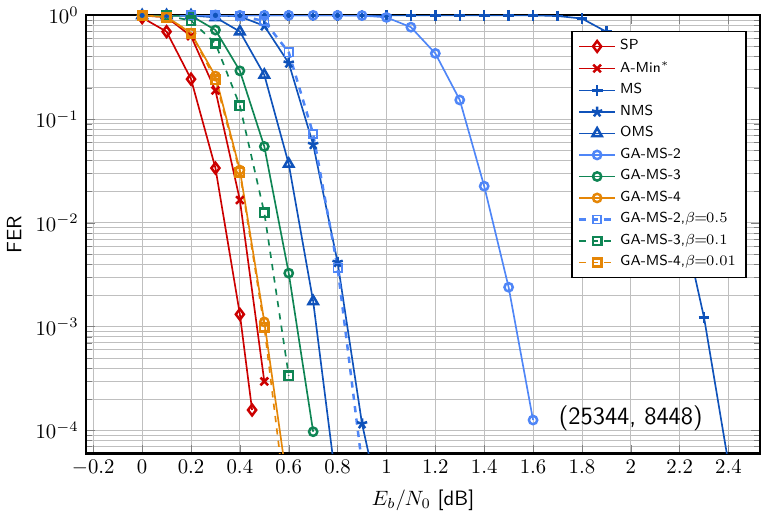}
	\caption{Floating-point FER comparison of SP, \amin, MS, NMS, OMS, and GA-MS decoding with $\beta$ and $\gamma\in\{2,3,4\}$ for 5G~NR LDPC codes (BG1, $R=\frac{1}{3}$, $Z=384$, and $K_u=22$) using QPSK and $I_{\max}=15$.}
	\label{fig:fer_flp_GAMS}
\end{figure}

The floating-point FER performance comparison of SP decoding and our GA-MS decoding with varying values of $\gamma$ and $\beta$ is provided in Fig.~\ref{fig:fer_flp_GAMS}.
The simulation is conducted with 5G~NR LDPC codes (BG1, $R=\frac{1}{3}$, $Z=384$, $K_u=22$) with quadrature phase shift keying (QPSK) over an additive white Gaussian noise (AWGN) channel and $I_{\max}=15$.
To fully demonstrate the capabilities of OMS and NMS decoding~\cite{chen2005TCOM}, we finely adjust the offset and the normalization factor for their optimal performance.
Specifically, for each $E_b/N_0$ point, we sweep offsets of OMS within the range of $0.3$ to $0.7$, in increments of $0.05$, and normalization factors of NMS between $\frac{1}{2}$ and $\frac{15}{16}$, in $\frac{1}{16}$ increments to plot the FER for the best values.
This approach ensures a fair comparison with GA-MS decoding.
In all following captions referring to GA-MS decoding, the last digit of the label (e.g., GA-MS-2) denotes the number $\gamma$ of used minima during the decoding.
When $\gamma=2$, GA-MS decoding is simplified to a near-MS algorithm (GA-MS-2) that surpasses MS decoding by $0.75$~dB, but still underperforms OMS and NMS.
If we increase $\gamma$ to $3$ or $4$, GA-MS decoding enables a significant improvement, closing the gap with SP decoding to $0.25$~dB and $0.1$~dB at FER~$=10^{-3}$, respectively.
With respect to determining the compensation factor, a moderate subtraction of $\beta$ can cause a notable performance improvement when $\gamma\in\{2,3\}$.
For example, GA-MS-3 decoding with a value of $\beta$ equal to $0.1$ only has a gap compared to SP decoding by $0.16$~dB and outperforms OMS decoding by $0.15$~dB at FER~$=10^{-3}$.
However, as $\gamma$ increases, especially for $\gamma\geq 4$, the original outgoing R-message in~\eqref{eq:GA-MSbasic} is accurate enough to approach the original \amin result so that the impact of $\beta$ is rapidly reduced.
As shown in Fig.~\ref{fig:fer_flp_GAMS}, GA-MS-4 decoding without $\beta$ has nearly the same error-correcting performance as that with a compensation~factor $\beta=0.01$.
Besides, GA-MS-4 also exhibits almost the same performance as the original \amin~decoding, which shows that $\gamma=4$ is precise enough for the result of \amin~decoding in~CN~processing.
\begin{table*}[t]
	\tabcolsep  1mm
	\renewcommand{\arraystretch}{1.0}
	\scriptsize
	\centering
	\caption{Comparison of Computational Complexity incurred by various LDPC decoding algorithms.}
	\begin{tabular}{llllllll}
		\toprule
		Algorithms  & SP$^{\dagger}$~\cite{jones2003approximate} & \amin$^{\dagger}$~\cite{jones2003approximate} & MS~\cite{fossorier1999reduced} & OMS~\cite{chen2005TCOM} & NMS$^{\ddagger}$~\cite{chen2005TCOM} & A-MS~\cite{Adj2017OPatent} & \textbf{GA-MS} 
		\\ \midrule
		Comparisons     & -                            & $(\bdc-1)\cdot M$    & $(2\bdc-3)\cdot M$ & $(2\bdc-3)\cdot M$         & $(2\bdc-3)\cdot M$ & $(2\bdc-3)\cdot M$              & $\bm{(\gamma\cdot\bdc-\frac{(\gamma+1)\cdot \gamma}{2})\cdot M}$
		\\ 
		Additions       & $\bdv\!\cdot\!N+(2\bdc\!-\!1)\!\cdot\!M$  & $\bdv\!\cdot\!N+(2\bdc\!-\!1)\!\cdot\!M$        & $\bdv\cdot N$      & $\bdv\cdot N+2M$  & $\bdv\cdot N + 2M$      & $\bdv\!\cdot\!N+(3\bdc\!-\!6)\!\cdot\!M$ & $\bm{\bdv\cdot N +2M}$                          
		\\ 
		LUTs    & $2\bdc\cdot M$                 & $(\bdc-1)\cdot M$    & -                & -                        & -                & $(2\bdc-4)\cdot M$              & $\bm{(\gamma-1)\cdot M}$                         
		\\ 
		Memory   & $\bdc\cdot M+N$               & $3M+N$               & $2M+N$             & $2M+N$                     & $2M+N$             & $3M+N$                          & $\bm{2M+N}$                                  
		\\ \bottomrule
	\end{tabular}			
	\label{tab:complexity}
	\begin{tablenotes}
		\footnotesize
		\item[*] $^{\dagger}$ The hyperbolic-tangent functions in CN processing for SP, \amin, and A-MS are approximated by LUTs in~\cite{jones2003approximate, Adj2017OPatent}, which refer to Table I of~\cite{kang2020enhanced}.
		\item[*] $^{\ddagger}$ The multiplication in NMS decoding is implemented by shift-and-add operations to reduce the computational complexity, similarly to~\cite{Su2022SSCL}.
	\end{tablenotes}
\end{table*}
\begin{figure}[t]
	\centering
	\includegraphics[width=0.95\columnwidth]{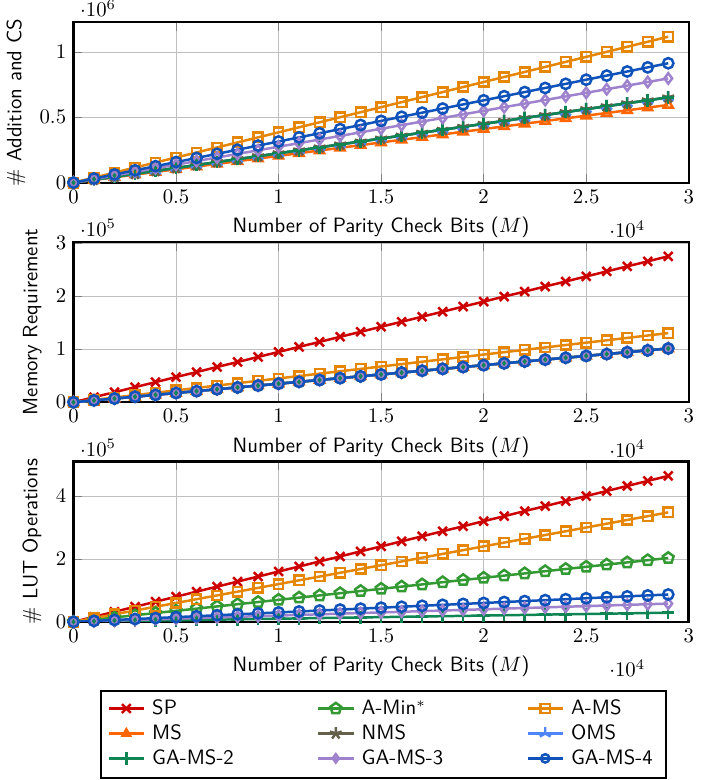}
	\caption{Comparison of computational complexity (addition and CS, memory requirement, and LUT operations) among GA-MS decoding and various existing LDPC decoding algorithms.}
	\label{fig:computationalComplexity}
\end{figure}

Table~\ref{tab:complexity} presents a summary of our analysis on the computational complexity per iteration for the proposed GA-MS decoding algorithm, in comparison with various other LDPC decoding algorithms~\cite{fossorier1999reduced, chen2005TCOM, jones2003approximate, Adj2017OPatent,Su2022SSCL}.
The analysis reveals that the number of comparisons in GA-MS decoding increases with $\gamma$ due to the internal sorting corresponding to multiple minima.
Nevertheless, only a single addition operation occurs outside the entire box-plus operator (as shown in~\eqref{eq:GA-MSadvanced}) within each CN processing unit.
As a result, GA-MS decoding outperforms the benchmark A-MS decoding in terms of both the number of addition and compare-select (CS) operations and the number of LUTs, especially when $\gamma\in\{2,3,4\}$.
GA-MS decoding also has the same memory consumption as MS-based decoding~\cite{fossorier1999reduced, chen2005TCOM}, which involves storing $N$ channel LLRs and two outgoing messages corresponding to~each~CN.

To further demonstrate the advantages of GA-MS decoding, we provide comparative plots in Fig.~\ref{fig:computationalComplexity}.
These plots chart a function of the parity-check length~$M$ (represented on the horizontal axis) for regular LDPC codes ($R=\frac{1}{3}$, $d_{c}=8$, $d_{v}=5$), to be consistent with the example in~\cite{verma2022low}.
Considering the maximum code length of 5G~NR LDPC codes, specified for $M=17664$, GA-MS-3 decoding exhibits a reduction of $28.7\%$ addition and CS operations and $22.3\%$ memory overhead reduction compared to \text{A-MS}~decoding.
It is noteworthy that GA-MS decoding requires slightly more LUT operations relative to OMS and NMS decoding.
Despite this, based on the message truncation in the CN processing, GA-MS-3 decoding reduces $87.5\%$ and $83.3\%$ of LUT operations compared to SP and \text{A-MS} decoding, respectively.

In conclusion, the proposed GA-MS decoding offers several distinct advantages over previous works.
First, by adjusting the parameter $\gamma$ to truncate the number of incoming messages, GA-MS decoding achieves a good trade-off between error-correcting performance and computational complexity.
Second, even with an increment of $\gamma$, GA-MS decoding still only computes two distinct outgoing messages using~\eqref{eq:GA-MSadvanced}, with no additional memory overhead.
In terms of hardware implementation, similar to an MS-based architecture, GA-MS decoding can completely decouple the procedures of the minima collection and approximate box-plus operators to maintain a high operating frequency, which is further~discussed~in~Section~\ref{sec:SecIV_all}.

\subsection{Quantization of GA-MS Decoding}\label{sec:SecIII-B}
In this section, we focus on quantization techniques to enhance the fixed-point performance of GA-MS decoding.
We adopt a uniform quantization:
\begin{equation}\label{eq:UniformQuan}
	\bm{\Lambda}(y_{v})=\sgn(y_{v})\cdot\min\left(\left\lfloor\frac{|y_{v}|}{\delta}+0.5 \right\rfloor, 2^{B-1}-1\right),
\end{equation}
where the bit-width $B$ is composed of one sign bit, $B_{\mathrm{i}}$ integer bits, and $B_{\mathrm{f}}$ fractional bits.
The channel gain factor $\delta$ is defined as ${1}/{2^{B_{\mathrm{f}}}}$ to scale the inputs.
During the LDPC decoding process, all propagating messages can be categorized into two types: $q_{v}$ and $t_{v}$ (associated with VNs), and $r_{c,v}$ (associated with CNs).
As $r_{c,v}$ is only based on the minimum of incoming T-messages, unlike an aggregate LLR, it has a smaller dynamic range.
Namely, $r_{c,v}$ can utilize fewer quantization bits than $q_{v}$ and $t_{v}$.
Therefore, we adopt a quantization scheme denoted as $(B_{\mathrm{VN}},B_{\mathrm{CN}},B_{\mathrm{f}})$, where $B_{\mathrm{VN}}$ and $B_{\mathrm{CN}}$ are the numbers of quantization bits for messages associated with VNs and CNs, respectively and all messages have $B_{\mathrm{f}}$ fractional~bits.
\begin{algorithm}[t]
	\small
	\SetKwInOut{Input}{Input}
	\SetKwInOut{Output}{Output}
	\SetKwInOut{Initialize}{Initialize}
	\SetKwInOut{Return}{Return}
	\caption{\texttt{Quantized~GA-MS-$\gamma$~Decoding (QC-Layered Version)}}
	\label{alg:gams_qc_layered}
	\Initialize{$\bm{q}_{v}\leftarrow\bm{y}_{v}$, $\bm{r}_{c,v}\leftarrow\mathbf{0}_{Z\times1},\forall c, v$}
	\vspace{+0.2cm}
	\tcp{Iterative decoding}
	\For{\upshape $i=0\;\textbf{to}\;I_{\max}-1$}
	{
		\For{\upshape $c=0\;\textbf{to}\;M_{p}-1$}
		{	
			\tcp{Initialize $\gamma$ minima and sign bit}
			$\mathbb{M}=[\bm{m}_1,\bm{m}_2,\hdots,\bm{m}_{\gamma}]\leftarrow\infty\cdot[\mathbf{I}_{Z\times1},\mathbf{I}_{Z\times1},\hdots,\mathbf{I}_{Z\times1}]$
			
			$\bm{s}\leftarrow\mathbf{I}_{Z\times 1}$
			
			\vspace{+0.2cm}
			\tcp{Phase 1: MIN}
			\For{\upshape $v\in\mathbb{V}_c$}
			{
				$\omega_v=\mathbf{H}_{p}[c][v],\;$ $\bm{t}_{v}\leftarrow$\textsf{\footnotesize cyclicShift}$(\bm{q}_{v}, \omega_v)-\bm{r}_{c,v}$
				
				\tcp{Sort~$\gamma$~minima,~find~index~of~$\bm{m}_{1}$}
				$[\mathbb{M},\bm{v}_{\min}]\leftarrow$\textsf{\footnotesize sortMin}$(\mathbb{M},|\bm{t}_{v}|)$
				
				$\bm{s}\leftarrow \bm{s}\cdot\sgn(\bm{t}_{v})$
			}
			
			\vspace{+0.2cm}
			\tcp{Phase 2: SEL}
			\For{\upshape $v\in\mathbb{V}_c$}
			{	
				\tcp{LUT-based~approx~using~$\gamma$~minima}
				$\bm{m}_{\LUT}\leftarrow$\textsf{\footnotesize LUTMin}$(\mathbb{M},\bm{v}_{\min}, v)$
				
				\tcp{update R- and Q-messages}
				$\bm{r}_{c,v}\leftarrow\bm{s}\cdot\sgn(\bm{t}_{v})\cdot\bm{m}_{\LUT}$
				
				$\bm{q}_{v}\leftarrow$\textsf{\footnotesize cyclicShift}$(\bm{t}_{v}+\bm{r}_{c,v},Z-\omega_{v})$
			}
		}
	}
	\Return{$\bm{\hat{x}}_v=\HD(\bm{q}_{v}),\forall v$}
\end{algorithm}
\begin{algorithm}[t]
	\small
	\caption{\texttt{LUTMin()}}
	\label{alg:gams_lut}
	\For{\upshape $k=0\;\textbf{to}\;Z-1$}
	{
		\If{$v\neq\bm{v}_{\min}[k]$} 
		{
			$\bm{m}_{\LUT}[k]=\bm{m}_{1}[k]$~\tcp{non-critical messages}
			
			\For{\upshape $t=2\;\textbf{to}\;\gamma$}
			{
				$\bm{m}_{\LUT}[k]=\textsf{\footnotesize LUT}(\bm{m}_{\LUT}[k],\bm{m}_{t}[k])$
			}
		}
		\Else{
			$\bm{m}_{\LUT}[k]=\bm{m}_{2}[k]$~\tcp{critical message}
			
			\For{\upshape $t=3\;\textbf{to}\;\gamma$}
			{
				$\bm{m}_{\LUT}[k]=\textsf{\footnotesize LUT}(\bm{m}_{\LUT}[k],\bm{m}_{t}[k])$
			}
		}
	}
\end{algorithm}

\subsubsection{Decoding Algorithm}
Quantized GA-MS-$\gamma$ decoding (for layered decoding of QC-LDPC codes) is outlined in Algorithm~\ref{alg:gams_qc_layered}.
As mentioned in Section~\ref{ref:Sec2_layered}, our GA-MS decoding requires memory for three message types ($\bm{q}_v[k]$, $\bm{t}_v[k]$, $\bm{r}_{c,v}[k]$), which are all $Z$-dimensional vectors for $0\leq v<N_p$, $0\leq c<M_p$, and $0\leq k<Z$.
Similar to~\cite{Studer2008Asilomar}, we divide the entire algorithm (excluding early termination) into two primary phases (MIN and SEL) to decouple the computation of~\eqref{eq:layeredMS} into several procedures that can be executed separately in different clock cycles.
This division is beneficial for hardware implementation to improve the maximum operating frequency.
First, we initialize the vectors $\bm{q}_{v}$ using the input channel LLR vectors $\bm{y}_{v}$ and set the vectors $\bm{r}_{c,v}$ to all-zero vectors.
The algorithm proceeds in a layer-wise manner (for each row of $\mathbf{H}_{p}$), and the message computation is executed iteratively only for the columns $v\in \mathbb{V}_{c}$ of $\mathbf{H}_{p}$ for which~$\mathbf{H}_{p}[c][v]\neq -1$.
Note that, in line~7, the magnitude set~$\mathbb{M}$ only contains the magnitude information of collected minima, and we need an extra vector $\bm{s}$ to store all sign bits. 

The first phase, referred to as MIN, calculates the intermediate vector $\bm{t}_{v}$ and gathers $\gamma$ minima vectors for the $\bm{t}_{v}$ in the current layer.
As a new vector $\bm{t}_{v}$ arises, the \textsf{\small sortMin()} function updates and maintains an ascending ordered set of $\gamma$ minima vectors, denoted as $\mathbb{M}=\{\bm{m}_{1},\bm{m}_{2},\hdots,\bm{m}_{\gamma}\}$, where $\bm{m}_{1}$ (a length-$Z$ vector) contains the first minima of $Z$ rows, $\bm{m}_{2}$ holds the second minima of $Z$ rows, and so on.
In addition, the \textsf{\small sortMin()} function also keeps track of the column index vector $\bm{v}_{\min}$ of $\bm{m}_{1}$ to distinguish the calculation of critical and non-critical messages~in~\eqref{eq:GA-MSbasic}.
\begin{figure}[t]
	\centering
	\begin{minipage}{.4\linewidth}
		\centering
		\includegraphics[width=\linewidth]{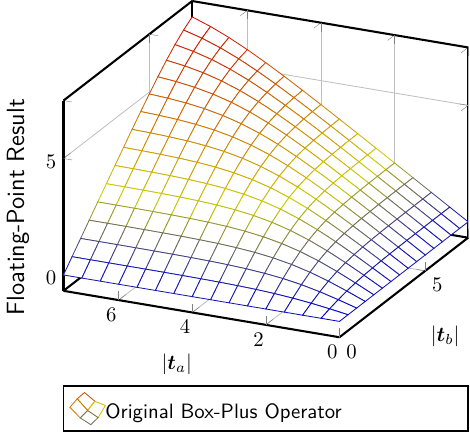}
		\label{fig:LUT0}
	\end{minipage}
	\quad
	\begin{minipage}{.4\linewidth}
		\centering
		\includegraphics[width=\linewidth]{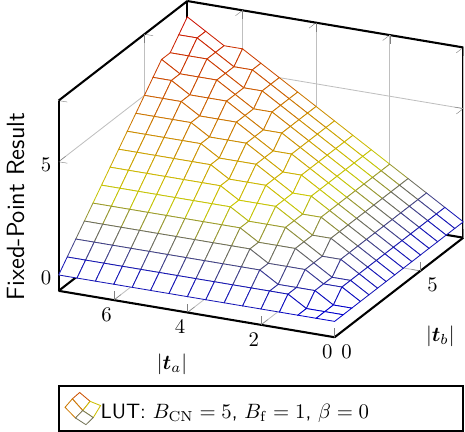}
		\label{fig:LUT1}
	\end{minipage}
	\quad
	\begin{minipage}{.4\linewidth}
		\centering
		\includegraphics[width=\linewidth]{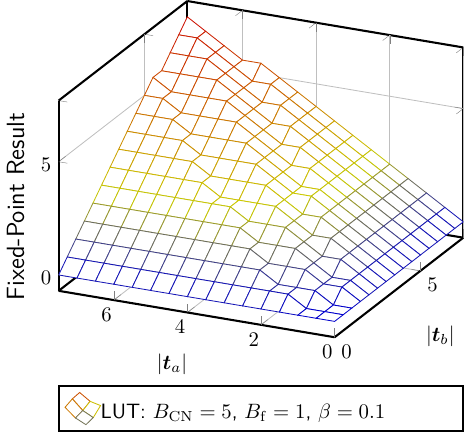}
		\label{fig:LUT2}
	\end{minipage}
	\quad
	\begin{minipage}{.4\linewidth}
		\centering
		\includegraphics[width=\linewidth]{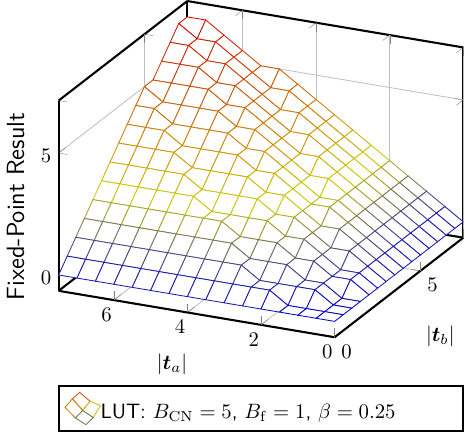}
		\label{fig:LUT3}
	\end{minipage}
	\caption{Comparison of the results between the original box-plus operator and LUTs with varying values of $\beta$ in~\eqref{eq:LUTGenerator}.}\label{fig:GA-MS-LUT}
\end{figure}
\begin{figure}[t]
	\centering
	\includegraphics[width=0.95\columnwidth]{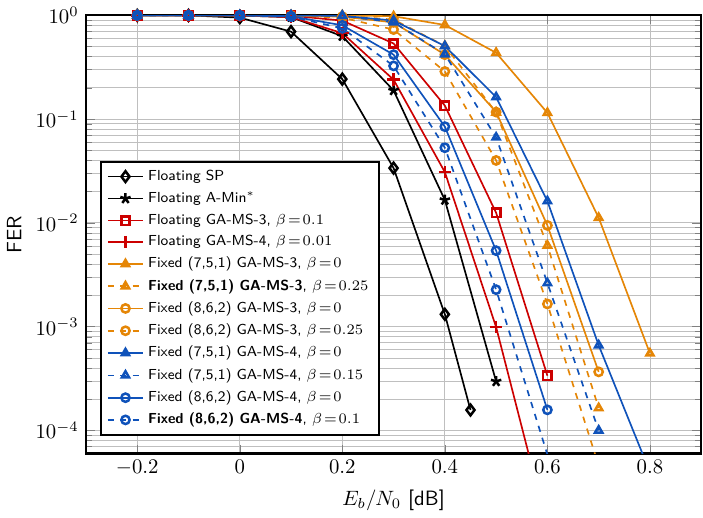}
	\caption{Fixed-point FER comparison of GA-MS decoding with different quantization strategies and varying values of $\beta$ for 5G~NR LDPC code (BG1, $R=\frac{1}{3}$, $Z=384$, and $K_u=22$) using QPSK and $I_{\max}=15$.}
	\label{fig:fer_fix_GAMS}
\end{figure}

The second phase, referred to as SEL, is activated after the MIN phase has swept across the entire layer.
During the SEL phase, the vectors $\bm{r}_{c,v}$ and $\bm{q}_{v}$ of the current iteration are block-wise updated based on the collected vectors $\mathbb{M}$, $\bm{v}_{\min}$, and $\bm{s}$.
The \textsf{\small LUTMin()} function shown in Algorithm~\ref{alg:gams_lut} corresponds to~\eqref{eq:GA-MSbasic} in Section~\ref{sec:SecIII_A}.
This function represents the recursive processing that repeatedly invokes the box-plus LUT at lines~$5$ and~$9$ (the efficient LUT design is discussed below), and the only difference is the removal of the vector $\bm{m}_{1}$ from the calculation if the current edge with the minimum incoming message.
As the new vector $\bm{m}_{\LUT}$ only contains the magnitude information, the updated vector $\bm{r}_{c,v}$ is computed by using both the sign vector $\bm{s}$ and the vector $\bm{m}_{\LUT}$.
Finally, the updated vector $\bm{q}_{v}$ is cyclically shifted using the inverse rotation $Z-w_{v}$.
Note that this additional rotation for Q-messages can be eliminated in hardware~\cite{Roth2010ASSCC}, as discussed~in~Section~\ref{sec:SecIV_highlevel}.

\subsubsection{LUT Design}
The box-plus LUT design is critical in determining error-correcting performance and computational complexity of GA-MS decoding.
The truncation of incoming messages to $\gamma$ minima, which are dominant in~CN~processing, brings advantages for GA-MS decoding, as it helps to design LUTs to approximate the original result.

If $t_{a}$ and $t_{b}$ have been quantized using the $(B_{\mathrm{VN}},B_{\mathrm{CN}},B_{\mathrm{f}})$ scheme, the non-linear term in~\eqref{eq:boxplus} can be quantized~as
\begin{equation}\label{eq:LUTnonlinear}
	\begin{aligned}
		\bm{\Lambda}(\triangle(t_{a}, t_{b}))=\sgn\left(\triangle(t_{a}, t_{b})\right)\!\cdot\!\left\lfloor\frac{\left|\triangle(t_{a}, t_{b})\right|}{\delta}\!+\!0.5\right\rfloor.
	\end{aligned}
\end{equation}
To realize a similar effect as the subtraction of a compensation factor in~\eqref{eq:GA-MSadvanced}, we move the $\beta$ into~\eqref{eq:LUTnonlinear} to introduce~\eqref{eq:LUTGenerator} as the approximation of the entire box-plus operator,
\begin{equation}\label{eq:LUTGenerator}
	\begin{aligned}
		&\LUT(t_{a}, t_{b})=\sgn(t_{a})\cdot\sgn(t_{b})\cdot\\
		&\left(\max\!\left(\!\min(|t_{a}|,\!|t_{b}|)-\!\left\lfloor\frac{\left|\triangle(t_{a},\! t_{b})\right|}{\delta}+\beta+0.5\right\rfloor,0\right)\!\right)\!.
	\end{aligned}
\end{equation}
\begin{figure*}[t]
	\centering
	\includegraphics[width=\linewidth]{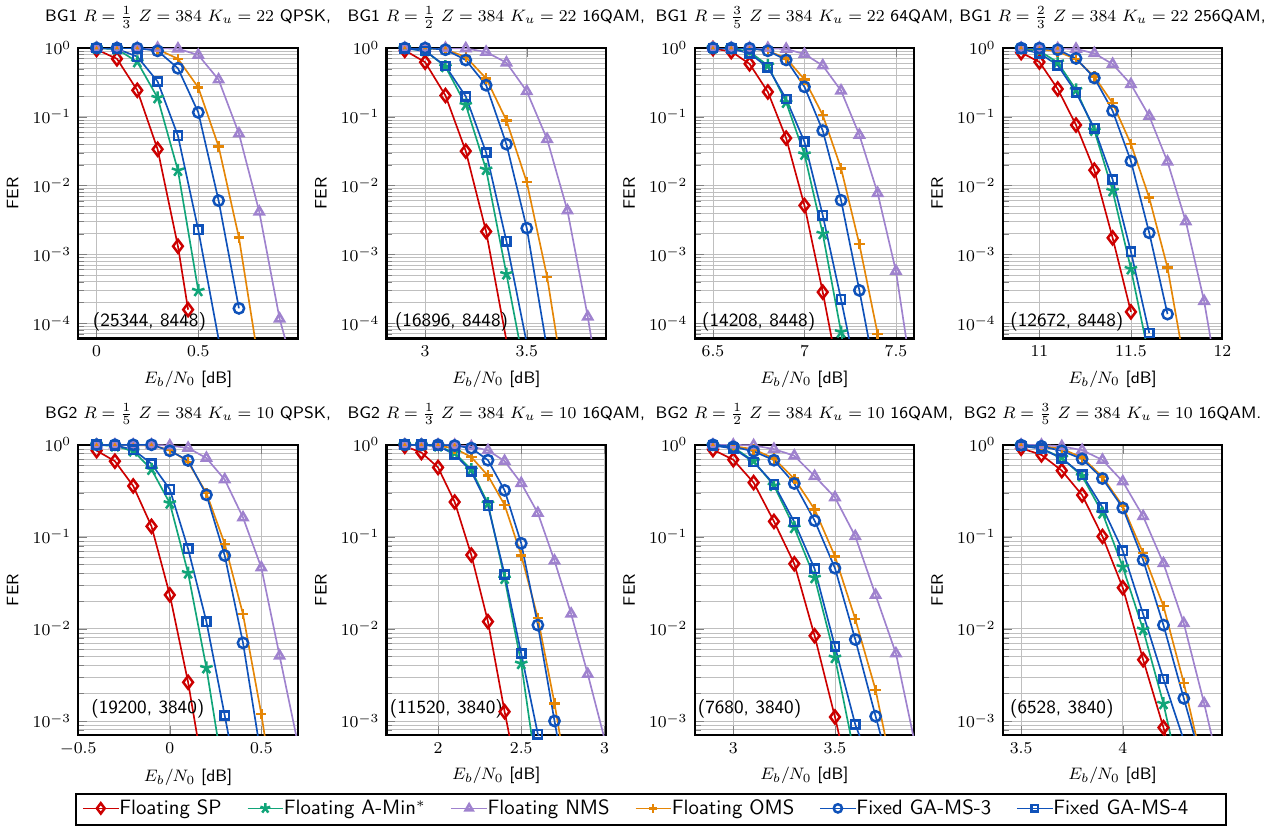}
	\caption{Fixed-point FER of GA-MS decoding for various 5G~NR LDPC codes and high-order modulations over an AWGN channel, where $I_{\max}=15$ and the parameters of OMS and NMS decoding have been carefully tuned.}
	\label{fig:MQAM}
\end{figure*}

Fig.~\ref{fig:GA-MS-LUT} demonstrates the impact of $\beta$ in~\eqref{eq:LUTGenerator} on the LUT results.
As $\beta$ gets larger, the difference between the LUT outcome and the original box-plus operator grows.
However, for two fully distinct magnitudes, the LUT result directly equals the minimum, corresponding to~\eqref{eq:boxplusMS}.
In general, if $\gamma$ is small, a larger $\beta$ is needed in~\eqref{eq:LUTGenerator} to compensate for performance degradation caused by truncation.
The parameter $\beta$ extends the design space of the LUTs for our GA-MS decoding, facilitating the exploration of various quantization strategies and high-order~modulations.
Based on a well-designed quantization strategy and $\beta$ in~\eqref{eq:LUTGenerator}, we can generate an efficient LUT and use~\eqref{eq:GA-MSbasic} as our fixed-point GA-MS decoding to approach the floating-point performance~of~\eqref{eq:GA-MSadvanced}.

\subsubsection{Comparison of Different Quantization Strategies}
Fig.~\ref{fig:fer_fix_GAMS} presents the fixed-pointed FER performance of GA-MS decoding under various quantization strategies and with different $\beta$ values in~\eqref{eq:LUTGenerator} (using the same code configuration as Fig.~\ref{fig:fer_flp_GAMS}).
First, we adopt two popular quantization schemes, $(7,5,1)$ and $(8,6,2)$, from the 5G~NR LDPC decoders in~\cite{cui2020design, lin202133Gbps, nadal2021parallel, Su2022SSCL, verma2022low}, to achieve a balance between high performance and hardware (including memory) complexity.
Second, we provide a selection of empirical $\beta$ values in~\eqref{eq:LUTGenerator} to improve quantized GA-MS decoding, which can be applied in the subsequent analysis and implementations.

Under the quantization schemes $(7,5,1)$ and $(8,6,2)$, fixed-point GA-MS-3 decoding with $\beta=0$ has a loss of $0.2$~dB and $0.1$ dB from floating-point GA-MS-3.
When using $\beta=0.25$ for $(7,5,1)$, the performance gap compared to floating-point GA-MS-3 reduces to $0.08$~dB.
However, GA-MS-4 decoding demands greater precision for message propagation.
The high-resolution $(8,6,2)$ scheme thus aligns well with GA-MS-4 decoding to enable more precise numerical calculation based on minima.
In Fig.~\ref{fig:fer_fix_GAMS}, the fixed-point performance using $(8,6,2)$ with $\beta=0.1$ almost has the same performance as that of floating-point GA-MS-4, which exhibits only a $0.1$ dB gap relative to floating-point SP decoding.
Consequently, for the following sections, our fixed-point GA-MS-3 decoding employs the quantization scheme $(7,5,1)$, while the fixed-point GA-MS-4 algorithm utilizes the quantization scheme~$(8,6,2)$.

\subsection{Comprehensive Performance Analysis}~\label{sec:SecIII-C}
To evaluate the error-correcting capability of the proposed GA-MS decoding in practical scenarios, we conduct simulations on various 5G~NR LDPC codes and with different high-order modulations~\cite{mao2016low}, comparing them with other classical LDPC decoding algorithms~\cite{Urbanke2001TIT, jones2003approximate, chen2005TCOM} for detailed performance analysis.
For all modulations, channel LLRs $y_{v}$ are obtained using the max-log-MAP method over an AWGN channel.
Notably, the configurable $\beta$ in~\eqref{eq:LUTGenerator} demonstrates robustness across a wide range of code rates and modulations, despite being relatively sensitive to the base graph selection.
Hence, we present empirical optimal values of $\beta$ in~\eqref{eq:LUTGenerator} for BG1 and BG2 in Fig.~\ref{fig:MQAM},~respectively.

For BG1, we set $\beta=0.25$ and $\beta=0.1$ for GA-MS-3 and GA-MS-4 decoding, respectively.
Numerical results show that fixed-point GA-MS-3 decoding can approach the performance of SP decoding within $0.25$ dB before FER~$=10^{-3}$, while fixed-point GA-MS-4 decoding exhibits a gap of $0.1$ dB compared to SP decoding.
This gap gradually diminishes as the code rate increases.
For BG2, we adopt a $\beta$ value of $0.1$ in~\eqref{eq:LUTGenerator} for both GA-MS-3 and GA-MS-4 decoding.
With $R=\frac{1}{5}$ with QPSK, fixed-point~GA-MS-4 decoding has an improvement of around $0.19$ dB compared to floating-point OMS decoding. 
Besides, at medium to high code rates, our fixed-point GA-MS-4 decoding performs almost the same as floating-point \amin~decoding on~both~BG1~and~BG2.

\section{Proposed 5G LDPC Decoder Architecture}\label{sec:SecIV_all}
In this section, we present the implementation of a fully reconfigurable 5G~NR LDPC decoder, incorporating our GA-MS decoding and all the aforementioned algorithmic optimizations.
This decoder is compatible with all 5G~NR LDPC codes.
We provide a comprehensive description of each core component, which  contains three embedded memory banks (referred to as the Q-memory, the T-memory, and the R-memory), a pool of node computation units (NCUs), a cyclic shifter unit (CSU), and a controller to complete layered GA-MS decoding algorithm in a block-parallel iteratively~decomposed~fashion.

\subsection{High-Level Overview}\label{sec:SecIV_highlevel}
\begin{figure}[t]
	\centering
	\includegraphics[width=0.9\columnwidth]{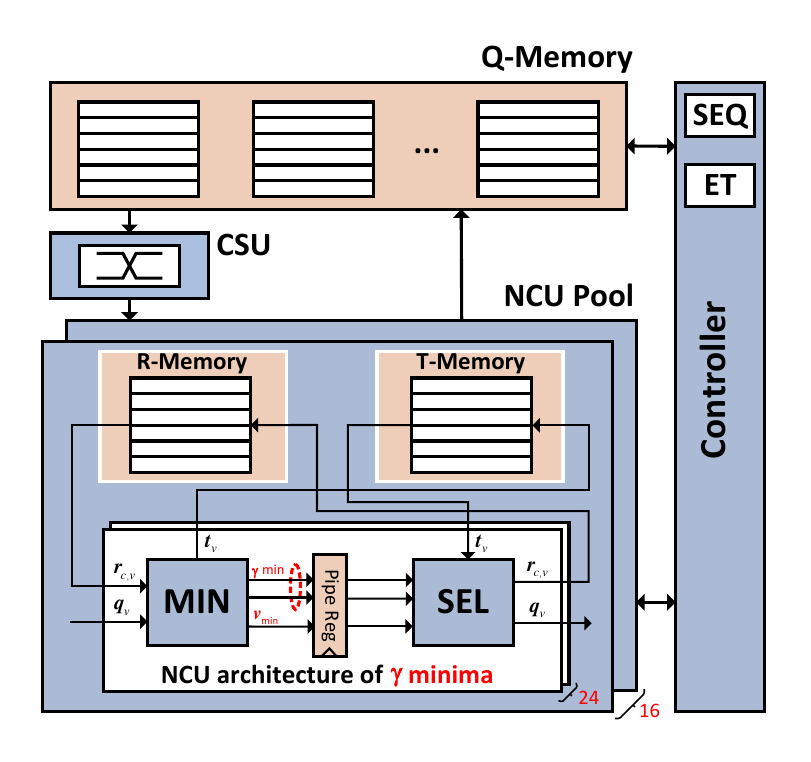}
	\caption{High-level architecture of fully reconfigurable 5G~NR LDPC decoder based on GA-MS decoding.}
	\label{fig:SecIV_Architecture}
\end{figure}

Fig.~\ref{fig:SecIV_Architecture} illustrates the high-level architecture of our fully reconfigurable 5G~NR LDPC decoder, which builds upon the baseline architecture proposed in~\cite{Studer2008Asilomar, Roth2010ASSCC}.
To enable layered GA-MS decoding in a block-parallel fashion, we decompose the processing of each layer into multiple cycles.
In each cycle, we update all $Z$ parity checks for the current block simultaneously (i.e., instantiating $Z_{\max}$ processing units for 5G) and further optimize the processing units by decoupling them into the MIN and SEL units to enhance the operating frequency.
As shown in Fig.~\ref{fig:SecIV_Architecture}, the MIN units perform several tasks in each cycle.
They read the corresponding vectors $\bm{q}_{v}$ and $\bm{r}_{c,v}$ from the associated Q- and R-memories, compute the intermediate vector $\bm{t}_{v}$, and write the results to the associated T-memory.
Additionally, the MIN units also track the set $\mathbb{M}$ with $\gamma$ minima vectors in Algorithm~\ref{alg:gams_qc_layered} and update pipeline registers of the NCUs at the end of each layer.
Meanwhile, based on the previously stored $\mathbb{M}$, the SEL units read the latest vector $\bm{t}_{v}$ from the associated T-memory and update the vectors $\bm{q}_{v}$ and $\bm{r}_{c,v}$.
It is noteworthy that the MIN and SEL units are pipelined to process two consecutive layers (the MIN units always work ahead of the SEL units).
Moreover, to rotate the Q-messages according to the QC-LDPC prototype matrix, we implement a CSU to perform a cyclic left-shift by $w=\mathbf{H}_{p}[c][v]$ for the read vector $\bm{q}_{v}$.
Instead of re-rotating the updated Q-messages when writing them back to the Q-memory, the rotation value of the Q-messages is tracked during the processing to avoid a second CSU instantiation~\cite{Roth2010ASSCC}.
Namely, in the hardware implementation, we remove the \textsf{\small cyclicShift()} function of line~$12$ and change line~$6$ to~\eqref{eq:SecIV_rerotation}~in~Algorithm~\ref{alg:gams_qc_layered}
\begin{equation}\label{eq:SecIV_rerotation}
	\omega_{v}=\!\!\!\!\!\mod(Z+\mathbf{H}_{p}[c][v]-\mathbf{H}_p[c-1\oplus M_{p}][v],\;Z),
\end{equation}
where the term $\mathbf{H}_p[c-1\oplus M_{p}][v]$ records the rotation value of the current block at the previous layer.

In our 5G~NR LDPC decoder, an early termination technique (same as~\cite{Studer2008Asilomar, Roth2010ASSCC}) called partial parity checks (PPCs) is adopted to terminate converged codewords and improve average throughput.
In the SEL units, each row of the prototype matrix yields a set of $Z$ parity checks which are combined into a single PPC.
If all PPCs are correct, the decoding procedure terminates prematurely after processing all $M_p$ rows of $\mathbf{H}_{p}$.
While the PPC approach is sub-optimal in terms of the number of decoding iterations compared to the conventional complete $\mathbf{H}\cdot\bm{\hat{x}}=0$, it can be done efficiently in a block-parallel~decoder.

Finally, the whole decoding process is orchestrated by the controller, which reads from a list of instructions in the SEQ memory and coordinates the components of our 5G~NR LDPC~decoder.
Notably, this controller can be configurable by the lifting size $Z$ (i.e., sub-block size), the prototype matrix $\mathbf{H}_{p}$, and $I_{\max}$, which enables~decoder~reconfigurability.

\subsection{Decoder Memories}\label{sec:SecIV_Memory}
In this section, we introduce decoder memories with grouping and data compression techniques to fit any 5G~NR LDPC codes into the allocated memories.
\begin{figure}[t]
	\centering
	\includegraphics[width=0.8\columnwidth]{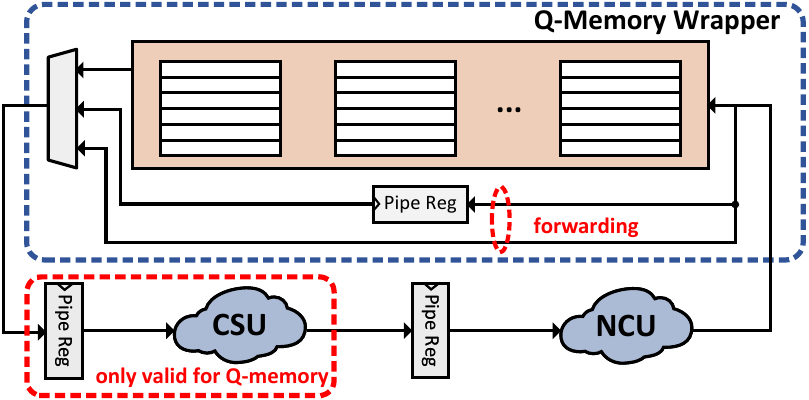}
	\caption{Memory wrapper for the Q-memories, which has the same structure as that of the T-memories.}
	\label{fig:SecIV_qtmem}
\end{figure}

\begin{table}[t]
	\tabcolsep  0.75mm
	\renewcommand{\arraystretch}{1.0}
	\def\CmidW{0.08cm}
	\centering
	\caption{Memory sizes of our 5G~NR LDPC decoder$^{\dag}$.}
	\begin{tabular}{llllll@{\hspace{1mm}}llll}
			\toprule
			& \multicolumn{4}{c}{$\bm{(7,5,1)}$}  & & \multicolumn{4}{c}{$\bm{(8,6,2)}$}                                                                   
			\\
			\textbf{Memory} & \multicolumn{1}{l}{Q} & \multicolumn{1}{l}{T} & \multicolumn{1}{l}{R-sign} & R-mag & & \multicolumn{1}{l}{Q} & \multicolumn{1}{l}{T} & \multicolumn{1}{l}{R-sign} & R-mag
			\\ \cmidrule(l{\CmidW}){2-5} \cmidrule(l{\CmidW}){7-10}
			\textbf{Width} [bit]    & \multicolumn{1}{l}{$168$}   & \multicolumn{1}{l}{$168$}   & \multicolumn{1}{l}{$24$}     & $312$  & & \multicolumn{1}{l}{$192$}   & \multicolumn{1}{l}{$192$}   & \multicolumn{1}{l}{$24$}     & $360$   
			\\
			\textbf{Depth} [word]    & \multicolumn{1}{l}{$68$}    & \multicolumn{1}{l}{$68$}    & \multicolumn{1}{l}{$316$}   & $46$  &  & \multicolumn{1}{l}{$68$}    & \multicolumn{1}{l}{$68$}    & \multicolumn{1}{l}{$316$}    & $46$   
			\\ \cmidrule(l{\CmidW}){2-5} \cmidrule(l{\CmidW}){7-10}
			\textbf{Instance Count} & $16$ & $16$ & $16$ & $16$ & & $16$ & $16$ & $16$ & $16$ \\
			\textbf{Capacity}$^{\ddagger}$~[KB] & $22.31$ & $22.31$ & $14.81$ & $28.03$ & & $25.5$ & $25.5$ & $14.81$ & $32.34$
			\\ \bottomrule
	\end{tabular}		
	\label{tab:memory}
	\begin{tablenotes}
		\item[*] $^{\dag}$ The number of all the above memory instances is $16$.
		\item[*] $^{\ddagger}$ Memory capacities are the sum of $16$ instances of each~memory.
	\end{tablenotes}
\end{table}

\subsubsection{Grouping}
As mentioned in Section~\ref{sec:SecII_5GLDPC}, 5G~NR LDPC codes feature $51$ distinct lifting sizes.
When $Z$ is less than $Z_{\max}$, it is inefficient and energy-consuming to consistently operate the decoder at maximum parallelism $Z_{\max}$.
Therefore, our LDPC decoder, including the NCU pool, datapaths, and memories, demands a fine-grained structure.

As illustrated in Fig.~\ref{fig:SecIV_Architecture}, we pack $24$ NCUs into a single group and thus divide the whole $Z_{\max}=384$ NCUs into $16$ groups that are driven by different gated clocks.
Each group shares a collection of independent memories to maintain macros with reasonable sizes and avoid extremely small (i.e., inefficient) macros.

Fig.~\ref{fig:SecIV_qtmem} depicts the corresponding memory wrapper for the Q-memories.
Each Q memory features a width of $24B_{\mathrm{VN}}$ bits and a depth of $68$ words, with $24$ denoting the number of Q-messages within each word, and $68$ corresponding to $N_{p}$ in BG1.
All the Q-memories share the same read and write addresses and thus merge a complete vector of length-$Z_{\max}$ messages to be read and written simultaneously.
As the same Q-memory block may undergo multiple updates during each iteration, we implement two forwarding paths to prevent possible memory conflicts and enhance throughput.
Notably, the depth of the T-memories can theoretically be reduced from $68$ to $19$ words, due to $d_{c}^{\max}=19$ in 5G base graphs.
However, this reduction necessitates a complicated peripheral circuit for address mapping~\cite{cui2020design}.
Furthermore, the memory area reduction for such small memories is less than proportional to the reduction in the number of words.
As a result, the T-memories in our decoder still maintain the same depth as the Q-memories~to~offer~simpler~control~logic. 

\subsubsection{Compressed Format of R-Messages}
In this decoder, we employ a compressed data format for R-messages instead of using explicit storage and the clipping of Q-messages like in~\cite{Studer2008Asilomar, Roth2010ASSCC}.
In~\eqref{eq:GA-MSbasic}, the outgoing R-message has only two distinct magnitudes in each row (i.e., critical and non-critical messages).
By excluding all sign bits, we can store a compressed word (only comprising two magnitudes and the column index of the critical message) to recover all R-messages for each row.
Hence, each group of the R-memories consists of two parts: R-sign and R-mag memories.
As mentioned in Section~\ref{sec:SecIII-B}, all sign bits of the R-messages demand explicit storage, the width of each R-sign memory is $24$ bits and the depth is $316$ words (i.e., maximum number of non-zero entries in BG1).
It is noticeable that the column index value can be compressed to require only $5$ bits (instead of the $7$ bits required to store the full column index), due to $d_{c}^{\max}=19$ in 5G.
In the decoder, we implement a LUT to perform this index compression operation.
For each R-mag~memory, the width is $24\times(2\times(B_{\mathrm{CN}}-1)+5)$~bits and the depth is $46$ words (corresponding to $M_{p}$ in BG1).
Compared to conventional explicit storage, the above compressed technique can save approximately $42.2\%$ and $46.9\%$ of bits for the R-memories with the $(7,5,1)$ and $(8,6,2)$ quantization schemes,~respectively.

In our 5G~NR LDPC decoder, the detailed memory configurations are outlined in Table~\ref{tab:memory}, with each memory instantiated $16$~times.
Hence, the total memory capacities for the two quantization schemes are $87.47$~KB for $(7,5,1)$ and $98.16$~KB~for~$(8,6,2)$.

\subsection{Node Computation Units (NCUs)}\label{sec:SecIV-C}
\begin{figure}[t]
	\centering
	\includegraphics[width=\columnwidth]{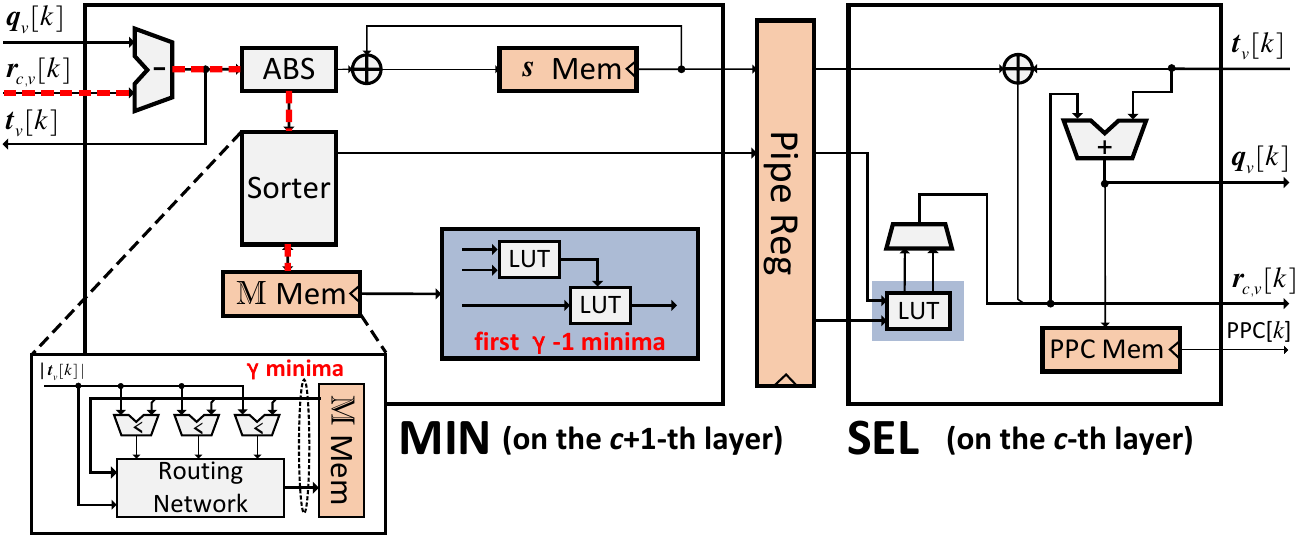}
	\caption{Architecture of the $k$-th NCU in the NCU pool for our GA-MS decoding, which illustrates the critical path (red dotted line) and internal sorter.}
	\label{fig:SecIV_gancu}
\end{figure}

The architecture of the $k$-th NCU ($0\leq k<Z_{\max}$) in the NCU pool is illustrated in Fig.~\ref{fig:SecIV_gancu}.
Internal pipeline registers separate the NCU computation into two phases.
The MIN unit iteratively computes the intermediate message $\bm{t}_{v}[k]$ and collects the updated $\gamma$ minima at the $((c+1)\cdot Z+k)$-th row, while the SEL unit concurrently updates the corresponding $\bm{q}_{v}[k]$ and $\bm{r}_{c,v}[k]$ at the $(c\cdot Z+k)$-th row.
Compared to the original NCU for layered OMS decoding~\cite{Studer2008Asilomar, Roth2010ASSCC} (only a simple subtraction with a fixed offset in the SEL unit), our GA-MS decoding in~\eqref{eq:GA-MSbasic} needs a set of LUTs, as shown in Algorithm~\ref{alg:gams_lut}.
This additional computation introduces latency in the SEL unit which degrades the maximum operating frequency of the decoder.
To alleviate this issue, we further decouple the partial calculation of~\eqref{eq:GA-MSbasic} and the updating of $\bm{q}_{v}[k]$ and $\bm{r}_{c,v}[k]$ into different cycles~to~balance~the~datapaths.
\begin{figure*}[t]
	\centering
	\includegraphics[width=0.95\linewidth]{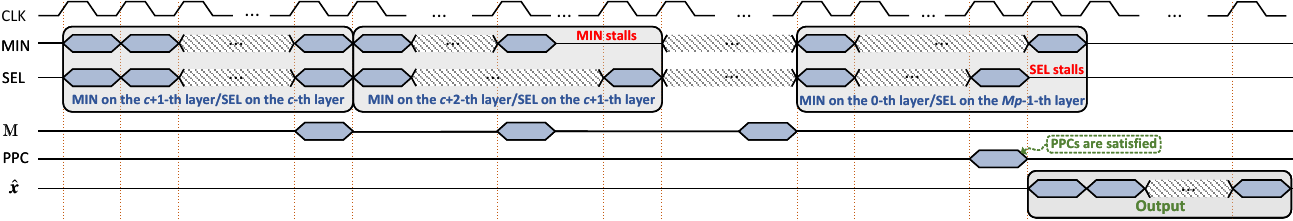}
	\caption{Example timing schedule of the proposed 5G~NR LDPC decoder based on GA-MS decoding.}
	\label{fig:SecIV_Timing}
\end{figure*}

For instance, when processing the non-critical message of~\eqref{eq:GA-MSbasic}, we need to sequentially invoke $\gamma-1$ LUTs for $\gamma$ minima inputs to calculate the result.
Before the MIN unit reaches the last block of each row, the iteratively updated $\mathbb{M}$ memory has already gathered (at least) $\gamma-1$ correct minima.
Hence, we can move the calculation of $\gamma-2$ LUTs, based on the first $\gamma-1$ minima of the $\mathbb{M}$ memory, to the MIN unit in advance.
Note that this result is only intermediate due to the absence of one minimum.
Upon arriving at the end of each row, the MIN unit forwards this intermediate result and $\gamma$ minima to pipeline registers.
The SEL unit only needs to perform one LUT based on the fully updated $\gamma$ minima to accurately compute the non-critical message.
This approach can significantly optimize the datapaths without any stalls.
The critical message of~\eqref{eq:GA-MSbasic} is processed similarly.
Moreover, due to the existing strict order of $\gamma$ minima (in the $\mathbb{M}$ memory), we can implement a pruned $\gamma+1\rightarrow\gamma$ sorter to eliminate the redundant comparators in the MIN unit.
This sorter, which comprises a routing network and a layer of $\gamma$ comparators as shown in Fig.~\ref{fig:SecIV_gancu}, can be considered as a special case of low-latency \emph{rank-order sorters}~\cite{le2020low}.

\subsection{Timing Schedule and Latency Analysis}\label{sec:SecIV_Timing}
Fig.~\ref{fig:SecIV_Timing} demonstrates the timing schedule of our 5G~NR LDPC decoder from the perspective of the NCUs.
As discussed before, the MIN and SEL units are pipelined to work on two consecutive layers to balance the datapaths.
However, this approach inevitably introduces stalls in the LDPC decoder.
In general, these stalls are categorized into two types: \circled{$1$} data dependency and \circled{$2$} row synchronization.
First, data dependency arises when the MIN units attempt to access a block for the Q-memories and T-memories, but the SEL units have not yet updated it.
Consequently, the MIN units must wait for the updated Q- and R-messages until the SEL units release this block.
Second, our LDPC decoder employs row synchronization to manage the decoding schedule and simplify the control logic, which is beneficial to decode 5G~NR LDPC codes with flexible code lengths and rates.
However, this synchronization results in additional stalls if two consecutive layers have different row degrees.
The decoding latency of our 5G~NR LDPC decoder is presented in~\eqref{eq:latency5GLDPC}, where the bound is the summation of non-zero entries, $I$ is the actual iteration number, and $\mathcal{D}_{c}$ is the delay of~\circled{$1$} at the $c$-th layer.
\begin{equation}\label{eq:latency5GLDPC}
	\mathcal{L} = I\cdot\left(\underbrace{\sum_{c=0}^{M_{p}-1}d_{c}}_{\mathrm{Bound}}+\!\!\!\!\!\!\underbrace{\sum_{c=0}^{M_p-1}\mathcal{D}_c}_{\mathrm{stalls\;from\;}\circled{$1$}}\!\!\!+\!\!\underbrace{\sum_{c=0}^{M_p-1}\!\!\max\left(d_{c-1\oplus M_p}-d_{c}, 0\right)}_{\mathrm{stalls\;from\;}\circled{$2$}}\right).\\
\end{equation}
\begin{figure}[t]
	\centering
	\includegraphics[width=0.95\columnwidth]{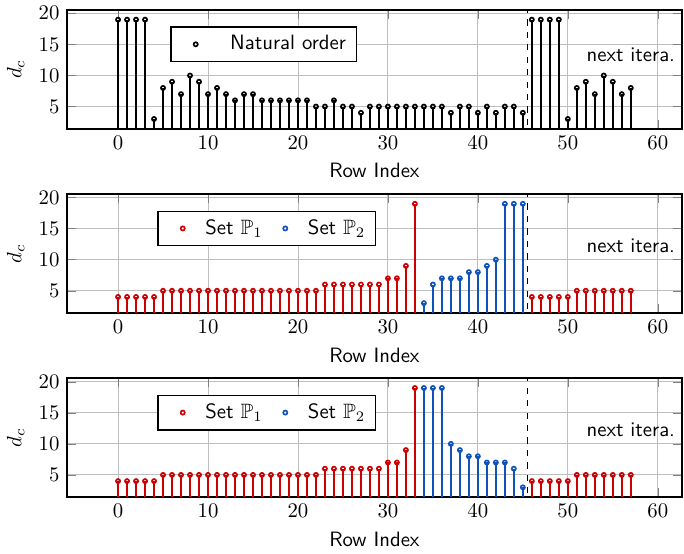}
	\caption{Row-degree distributions of BG1 using various static schedule techniques (natural order, OSS 1-2, OSS 1-3).}
	\label{fig:OSS1}
\end{figure}

\subsection{Optimized Static Schedule (OSS)}\label{sec:SecIII_OSS}
In the following, we will optimize the schedule of the decoder to improve convergence and to reduce the number of stall cycles.
Since the posterior LLRs are updated in a row-wise fashion, the convergence speed is greatly influenced by the order in which layers are processed.
Generally, this order can be determined by either dynamic schedules~\cite{Chang2021Dynamic} or static schedules~\cite{liang2019hardware,tian2022novel}.
Static schedules offer a computational complexity advantage over dynamic schedules, as they do not require real-time calculations.
Notably, some static schedule techniques are proposed in~\cite{liang2019hardware,tian2022novel} for 5G~NR LDPC codes, but they ignore potential impact on throughput due to hardware constraints.
In this section, we introduce a hardware-friendly OSS approach tailored to 5G~NR.
This OSS scheme delivers a $0.05$ dB performance gain compared to conventional layered decoding and reduces the worst-case latency by around~$15\times I_{\max}$~cycles, compared~to~natural~layer~ordering.

First, we adopt two classical optimization principles to improve the error-correcting performance.
In 5G, the first two columns of the base graphs are punctured to boost transmission efficiency.
Let $\mathbb{P}_{i}$, $i\in\{0,1,2\}$, denote the sets of row indices in the base graphs with zero, one, and two punctured non-zero entries, respectively.
The first optimization principle of our schedule dictates that we prioritize rows with fewer punctured non-zero entries.
Subsequently, for the rows in the same $\mathbb{P}_{i}$, we decode the rows with smaller $d_{c}$ first.
These two optimization principles (least punctured and least row-degree) are also used in the BG based static schedule (BGSS) in~\cite{tian2022novel} to speed up the~decoding~convergence.
\begin{figure}[t]
	\centering
	\includegraphics[width=0.95\columnwidth]{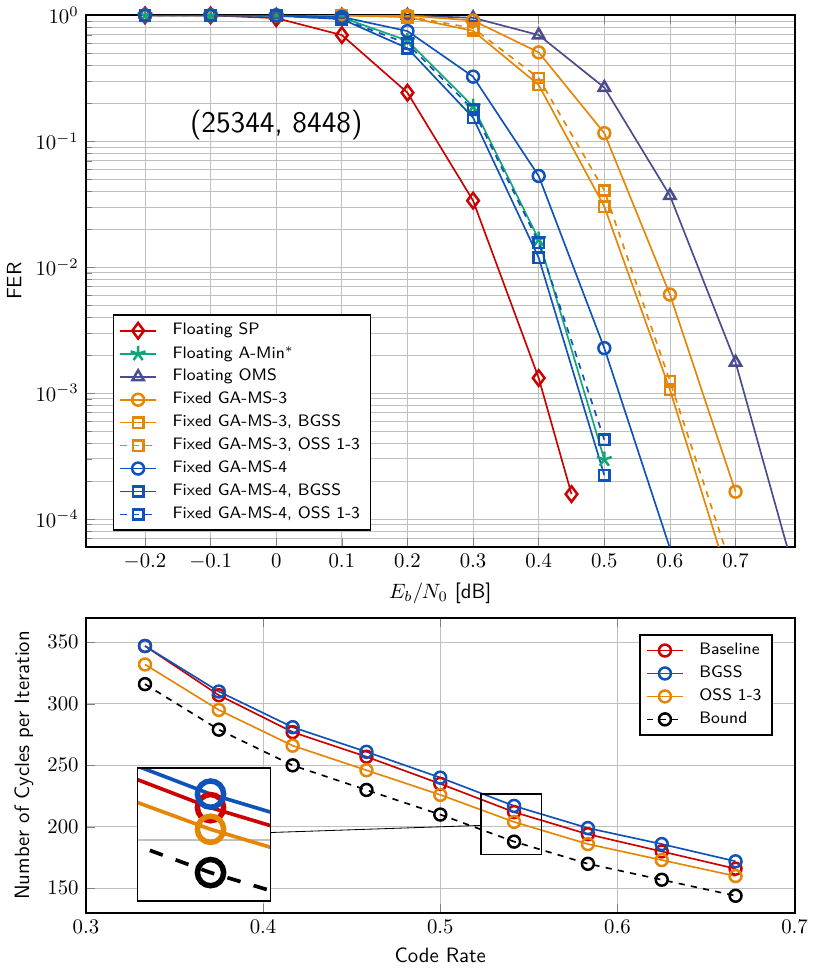}
	\caption{FER performance and latency analysis for the proposed OSS scheme on BG1.}
	\label{fig:oss_performance}
\end{figure}

The third optimization principle of our OSS scheme aims to diminish the worst-case latency in block-parallel architectures.
As outlined in Section~\ref{sec:SecIII-B}, classical LDPC block-parallel decoders~\cite{Studer2008Asilomar, Roth2010ASSCC} decouple~\eqref{eq:layeredMS} into several steps (e.g. the MIN and SEL phases) and execute them separately over different cycles.
In most cases, the aforementioned steps can nearly overlap at two consecutive rows in the base graphs, i.e., each row can be processed within $d_c$ cycles, which is also the latency bound of LDPC block-parallel decoders.
However, possible stalls occur when two consecutive rows share column indices (data dependency) or have apparently higher row degrees (row synchronization), which is explained in detail in Section~\ref{sec:SecIV_Timing}.
Especially for the latter, stalls are unavoidable due to starvation of the pipeline.
Hence, we arrange the rows of $\mathbb{P}_{2}$ in descending order of $d_{c}$ to ensure that adjacent rows have preferably similar row degrees.

Fig.~\ref{fig:OSS1} displays the row-degree distributions of BG1 using various static schedule techniques.
Note that the set $\mathbb{P}_{0}$ is empty in BG1.
Based on OSS 1-2 (incorporating the first two principles, equivalent to BGSS in~\cite{tian2022novel}), row-degree discontinuities appear at the junctions between $\mathbb{P}_{1}$ and $\mathbb{P}_2$, as well as between two consecutive iterations, resulting in redundant stalls in the LDPC decoder.
However, OSS 1-3 (adopting all three principles) balance the error-rate and decoding latency.
Since the majority of rows ($74\%$) still adhere to the least punctured and least row-degree principles, OSS 1-3 features a fast decoding convergence.
Then, by ordering the set $\mathbb{P}_{2}$ by descending $d_c$, we can minimize unnecessary stalls at the junctions of internal iterations.
Importantly, pruning the columns or adjusting code rates only extends evenly on both sides of the core rows (the first four rows with the maximum row-degree $d_c^{\max}$ in BG1 and BG2) and does not affect the property that adjacent rows have similar row degrees in OSS 1-3.
Consequently, the OSS algorithm is compatible with all 5G~NR LDPC codes.
\begin{table}[t]
	\scriptsize
	\tabcolsep 0.15mm
	\renewcommand{\arraystretch}{1.0}
	\def\CmidW{0.08cm}
	\caption{ASIC results of the proposed 5G~NR LDPC decoder.}
	\label{tab:5GSynthesisPostlayout}
	\centering
	\begin{tabular}{lcccccccc}
		\toprule
		& \multicolumn{8}{c}{\textbf{This Work}}                                                                                                                                                                                   \\ \cmidrule(l{\CmidW}){2-9}
		\textbf{Technology} {[}nm{]}         & \multicolumn{8}{c}{28}                                                                                                                                                                                          \\ 
		\textbf{Algorithm}                   & \multicolumn{8}{c}{GA-MS-3}                                   \\ 
		\textbf{Iterations}                  & \multicolumn{8}{c}{$4$} 
		\\ 
		\textbf{Voltage} {[}V{]}             & \multicolumn{8}{c}{$1.0$}                                                                     \\ \cmidrule(l{\CmidW}){2-9}
		\textbf{Implementation}              & \multicolumn{4}{c}{Synthesis}      &
		\multicolumn{4}{c}{Post-layout}                                                             \\
		\textbf{Core Area} {[}mm$^{2}${]}    &
		\multicolumn{4}{c}{$1.274$}         & \multicolumn{4}{c}{$1.823$}                                                                   \\ 
		\textbf{Frequency} {[}MHz{]}         &
		\multicolumn{4}{c}{$1250$}            & \multicolumn{4}{c}{$895$}                                                                     \\
		\textbf{T/P$^\dagger$} {[}Gbps{]}    & \multicolumn{1}{c}{$24.58^\wr$} & \multicolumn{1}{c}{$34.11^*$} & \multicolumn{1}{c}{$30.29^\star$} &
		\multicolumn{1}{c}{$34.00^+$} & \multicolumn{1}{c}{$17.60^\wr$} & \multicolumn{1}{c}{$24.42^*$} & \multicolumn{1}{c}{$21.69^\star$} &
		\multicolumn{1}{c}{$24.34^+$} \\ \cmidrule(l{\CmidW}){2-5} \cmidrule(l{\CmidW}){6-9}
		\parbox{1.8cm}{\raggedright \textbf{Area Eff.$^\dagger$}\\ {[}Gbps/mm$^{2}${]}} &
		\multicolumn{1}{c}{$19.29^\wr$} &  \multicolumn{1}{c}{$26.77^*$} & \multicolumn{1}{c}{$23.78^\star$} & \multicolumn{1}{c}{$26.69^+$} & \multicolumn{1}{c}{$9.66^\wr$} &  \multicolumn{1}{c}{$13.40^*$} & \multicolumn{1}{c}{$11.90^\star$} & \multicolumn{1}{c}{$13.36^+$} \\ \bottomrule
	\end{tabular}
	\begin{tablenotes}
		\item[*] $^\wr$ BG1, $R=\frac{1}{3}$. $^{*}$ BG1 $R=\frac{8}{9}$. $^{\star}$ BG2, $R=\frac{1}{5}$. $^{+}$ BG2, $R=\frac{2}{3}$.
		\item[*] $^\dagger$ We set a fixed number of iterations to $4$, without using early termination.
	\end{tablenotes}
\end{table}

Fig.~\ref{fig:oss_performance} illustrates that our OSS algorithm can yield a $0.05$~dB improvement at FER~$=10^{-3}$.
Fixed-point GA-MS-4 decoding with OSS only has a gap of $0.05$~dB compared to floating-point SP decoding, and even outperforms floating-point \amin~decoding before FER~$=10^{-3}$.
Furthermore, we evaluate the number of cycles required per iteration using various static schedule schemes in Fig.~\ref{fig:oss_performance}.
Our baseline is the conventional layered decoding~\cite{Studer2008Asilomar}.
The black dashed line represents the summation of non-zero entries in $\mathbf{H}_{p}$, serving as the lower bound on the number of cycles (no stalls) of a single iteration based on a block-parallel architecture.
It is obvious that despite a $0.05$~dB error-correcting improvement offered by BGSS, its row-degree discontinuities lead to increased stalls in a single iteration.
In contrast, our OSS approach can reduce around $15$ cycles per iteration, especially at low code rates, which is beneficial to alleviate the worst-case latency in practical communication scenarios.

Therefore, the aforementioned stalls in~\eqref{eq:latency5GLDPC} can mostly be avoided through reasonable column reordering and the proposed OSS scheme.
For stalls from~\circled{$1$}, they can be removed by a simple column reordering~\cite{lin202133Gbps}.
Specifically, in two consecutive layers, we allow the MIN units to first visit independent blocks and let the SEL units visit dependent blocks, which can make most of $\mathcal{D}_{c}$ equal to $0$, especially at low to medium code rates.
For stalls from~\circled{$2$}, following the least punctured and least row-degree principles, our OSS scheme can provide a layer reordering that has only one peak in the row-degree distribution.
Therefore, the latency of our 5G~NR LDPC decoder can be simplified as~\eqref{eq:latency5GLDPCadvanced} from~\eqref{eq:latency5GLDPC}
\begin{equation}\label{eq:latency5GLDPCadvanced}
	\mathcal{L} \approx I\cdot\left(\sum_{c=0}^{M_{p}-1}d_{c}+\left|d_{c}^{\max}-d_{c}^{\min}\right|\right).\\
\end{equation}

\section{Implementation Results}\label{sec:SecV_synthesis}
In this section, we present the implementation results of our 5G~NR LDPC decoder based on a STM $28$nm FD-SOI technology.
The decoder is synthesized by Synopsys Design Compiler and placed and routed using Cadence Innovus Implementation System.
Power analysis is done under typical operating conditions ($1.0$~V and $25~^\circ$C).
To balance error-correcting performance and hardware complexity, we employ the quantization scheme $(7,5,1)$ (as discussed in Section~\ref{sec:SecIII-B}) and incorporate fixed-point GA-MS-3 decoding into our decoder.
All memory macros are based on STM $28$nm FD-SOI dual-port SRAM. 
The size of the LUTs in the NCUs is $16\times16$ and each value in~\eqref{eq:LUTGenerator} is quantized as $4$ bits.
We instantiate $384$ NCUs in the NCU pool to support the maximum lifting size in 5G.
When $Z<Z_{\max}$, our decoder operates in a more fine-grained fashion by dividing into $16$ groups.
Each group is driven incrementally by different gating clocks, ensuring that block $i$ ($i=0,1,\hdots,15$) is activated only if all other blocks $j$ ($j=0,1,\hdots,i-1$) are active.
Based on the worst-case latency when using BG1 with $R=\frac{1}{3}$, the SEQ memory comprises $332$ instruction words (in line with~\eqref{eq:latency5GLDPCadvanced}), with each instruction being $59$ bits in size.

\begin{figure}[t]
	\centering
	\begin{minipage}{.475\linewidth}
		\centering
		\includegraphics[width=0.85\columnwidth]{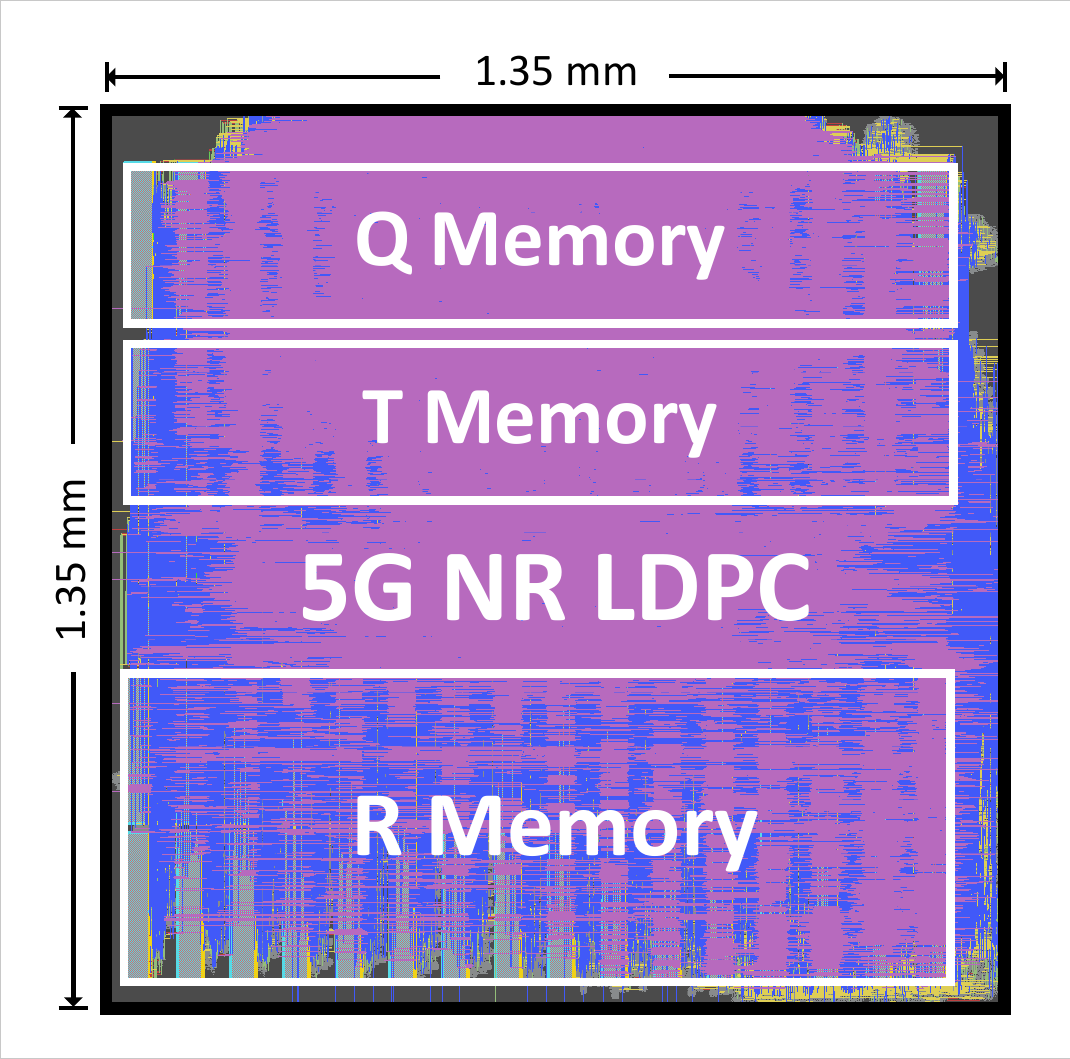}
	\end{minipage}
	\begin{minipage}{.475\linewidth}
			\centering
			\scriptsize
			\tabcolsep 1mm
			\renewcommand{\arraystretch}{1.125}
			\begin{tabular}{lc}
				\toprule
				\textbf{Technology}  & $28$nm FD-SOI \\ \midrule
				\multicolumn{1}{l|}{\textbf{Quantization} [bit]} & $(7,5,1)$ \\
				\multicolumn{1}{l|}{\textbf{Core Area} [mm$^{2}$]} & $1.35\!\times\!\!1.35$ \\
				\multicolumn{1}{l|}{\textbf{Gate Count} [M]} & $2.20$ \\ 
				\multicolumn{1}{l|}{\textbf{Voltage} [V]} & $1.0$ \\
				\multicolumn{1}{l|}{\textbf{Frequency} [MHz]} & $895$ \\
				\multicolumn{1}{l|}{\textbf{Peak T/P} [Gbps]} & $24.42$ \\
				\multicolumn{1}{l|}{\textbf{Power} [mW]} & $306.8$ \\
				\multicolumn{1}{l|}{\textbf{Energy} [pJ/bit]} & $12.56$ \\
				\bottomrule
			\end{tabular}
	\end{minipage}
	\caption{A post-layout of the proposed 5G~NR LDPC decoder implemented in a $28$nm process, wherein the white boxes represent the integrated Q-memory, T-memory, and R-memory, implemented by $28$nm FD-SOI dual-port SRAM.}
	\label{fig:SecV_postlayout}
\end{figure}

\begin{figure}[t]
	\centering
	\begin{minipage}{.45\linewidth}
		\centering
		\includegraphics[width=\columnwidth]{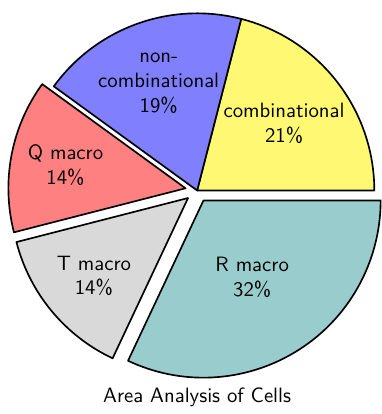}
	\end{minipage}
	\quad
	\begin{minipage}{.45\linewidth}
		\centering
		\includegraphics[width=\linewidth]{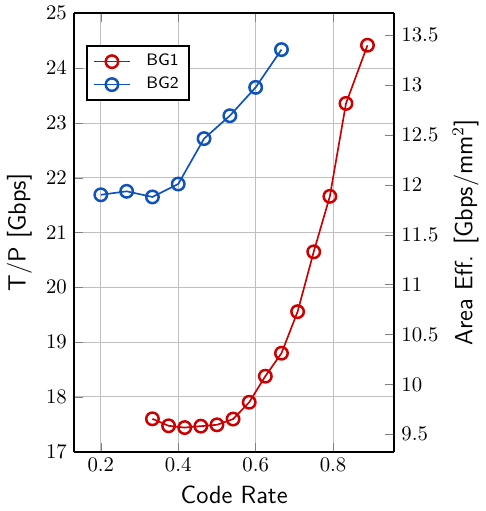}
	\end{minipage}
	\caption{Implementation results of our 5G~NR LDPC decoder in a $28$nm process across various code configurations and a fixed iteration value~of~$4$.}\label{fig:SecV_Plotresults}
\end{figure}
\begin{table}[t]
	\footnotesize
	\tabcolsep 2.5mm
	\renewcommand{\arraystretch}{0.95}
	\def\CmidW{0.08cm}
	\caption{Power breakdown$^\dagger$ of the proposed 5G~NR LDPC decoder.}
	\label{tab:PowerBreakdown}
	\centering
	\begin{tabular}{l c c}
		\toprule
		& \textbf{Power} {[}mW{]} & \textbf{Percentage} {[}\%{]} \\ \midrule
		Q-Memory           & $51.24$          & $16.7$           \\
	    T-Memory           & $43.93$          & $14.32$          \\
		R-Memory           & $82.5$           & $26.89$          \\
		CSU                & $35.13$          & $11.45$          \\
		NCU                & $32.52$          & $10.6$           \\
		Controller         & $22.83$          & $7.44$           \\
		IO Interfaces      & $6.65$           & $2.17$           \\
		Others             & $32.0$           & $10.43$          \\ \cmidrule(l{\CmidW}){2-3}
		Total              & $306.8$          & $100$            \\ \toprule
	\end{tabular}
	\begin{tablenotes}
	\item[*] $^\dagger$ We present the power breakdown in the case of $\Theta_{\mathrm{T/P}}$.
	\end{tablenotes}
\end{table}
\begin{table}[t]
	\scriptsize
	\centering
	\tabcolsep 1.1mm
	\renewcommand{\arraystretch}{0.95}
	\caption{Average T/P of proposed decoder with early termination.}
	\label{tab:AvgTP}
	\begin{tabular}{llllllll}
		\toprule
		$E_b/N_0$ {[}dB{]} & $3.0$  & $3.5$   & $4.0$   & $4.5$   & $5.0$   & $5.5$   & $6.0$   \\ \midrule
		\textbf{Avg Iter.}          & $15$   & $11.15$ & $6.05$  & $4.40$  & $3.49$  & $2.79$  & $2.35$  \\ 
		\textbf{Avg T/P} {[}Gbps{]} & $6.51$ & $8.76$  & $16.15$ & $22.20$ & $27.99$ & $35.01$ & $41.56$ \\ \bottomrule
	\end{tabular}
	\begin{tablenotes}
		\item[*] We use LDPC codes at BG1 with $R=\frac{8}{9}$ and $Z=384$.
	\end{tablenotes}
\end{table}

\subsection{Implementation Results for 5G~NR LDPC Codes}
\begin{table*}[]
	\scriptsize
	\tabcolsep 1.5mm
	\renewcommand{\arraystretch}{1.05}
	\def\CmidW{0.08cm}
	\caption{Comparisons with the state-of-the-art LDPC decoders.}
	\label{tab:5Gresults}
	\centering
	\begin{tabular}{lcccccccccc}
		\toprule
		& \multirow{2}{*}{\textbf{This work}} & TCAS-I'21 & ISCAS'21 & SSCL'22 & TCAS-II'22 & TVT'23 & TCAS-I'22 & ASSCC'10 & TVLSI'15 & JSSC'10 \\
		&    & \cite{cui2020design}$^\diamond$ & \cite{lin202133Gbps}$^\diamond$ & \cite{Su2022SSCL} & \cite{yun2021area}$^\diamond$ & \cite{verma2022low} & \cite{Lee2022TCASI} & \cite{Roth2010ASSCC} & \cite{kumawat2015high} & \cite{zhang2010JSSC}\\ \cmidrule(l{\CmidW}){2-2} \cmidrule(l{\CmidW}){3-3} \cmidrule(l{\CmidW}){4-4} \cmidrule(l{\CmidW}){5-5}
		\cmidrule(l{\CmidW}){6-6} \cmidrule(l{\CmidW}){7-7}
		\cmidrule(l{\CmidW}){8-8} \cmidrule(l{\CmidW}){9-9}
		\cmidrule(l{\CmidW}){10-10} \cmidrule(l{\CmidW}){11-11}
		\textbf{Technology} {[}nm{]}         & $28$                    & $90$         & $28$           & $40$         & $65$      & $90$         & $65$          & $90$      & $90$        & $65$             \\
		\textbf{Algorithm}                   & GA-MS-3                 & IAMS         & NMS            & NMS          & MS        & SOMS         & OMS           & OMS       & OMS         & OMS              \\
		\textbf{Implementation}              & Post-layout             & Synthesis    & Post-layout    & Silicon      & Synthesis & Post-layout  & Post-layout   & Silicon   & Post-layout & Silicon          \\
		\textbf{Voltage} {[}V{]}             & $1.0$                   & $-$          & $-$            & $0.9$        & $-$       & $1.0$        & $1.2$         & $1.0$     & $0.9$       & $0.7$            \\
		\textbf{Standard}                    & 5G~NR                   & 5G~NR        & 5G~NR          & 5G~NR        & 5G~NR     & 5G~NR        & 5G~NR         & 802.11n   & 802.11n     & 10GBASE-T        \\
		\textbf{Architecture}                & block          & row          & block          & row          & row       & row          & partial       & block     & block       & partial          \\ 
		\textbf{Iterations}                  & $4$                     & $15$         & $1$            & $5$          & $10$      & $10$         & $3$           & $10$      & $10$        & $8$              \\
		\textbf{Max Code Length}             & $26112$                 & $2600$       & $26112$        & $6400$       & $1664$    & $3072$       & $26112$       & $1944$    & $1944$      & $2048$           \\
		\textbf{Frequency} {[}MHz{]}         & $895$             & $158.2$      & $556$          & $180$        & $244$     & $192.3$      & $500$         & $346$     & $336$       & $100$            \\
		\textbf{Area} {[}mm$^{2}${]}           & $1.823$           & $1.353$      & $1.97$         & $2.07$       & $1.16$    & $6.45$       & $5.74$        & $1.77$    & $5.2$       & $5.35$           \\
		\textbf{Gate Count} {[}M{]}          & $2.20$             & $0.24$       & $2.83$         & $1.69$       & $0.81$    & $-$          & $2.67$        & $-$       & $0.51$      & $-$              \\
		\textbf{Peak T/P$^{*}$} {[}Gbps{]}              & $24.42$            & $0.914$      & $33.2$         & $2.29$       & $4.1$     & $9.6$        & $21.78$       & $0.679$   & $1.71$      & $2.13$           \\
		\textbf{Power} {[}mW{]}              & $306.8$            & $76.4$       & $232$          & $139.4$      & $115.8$   & $3456$       & $413$         & $107.3$   & $451.3$     & $144$            \\ \midrule
		\multicolumn{11}{l}{Scaled to $28$nm, $1.0$~V$^{\dagger}$, and a fixed iteration value of $4^{\ddagger}$}\\
		\textbf{Area} {[}mm$^{2}${]}           & $1.823$      & $\bm{0.131}$      & $1.97$         & $1.014$      & $0.215$  & $0.624$      & $1.065$        & $0.171$   & $0.503$     & $0.993$    \\
		\textbf{Peak T/P} {[}Gbps{]}              & $24.42$       & $11.02$       & $8.3$         & $4.09$       & $23.79$  & $\bm{77.14}$      & $37.92$        & $5.46$    & $13.74$     & $9.89$     \\
		\textbf{Area Eff.} {[}Gbps/mm$^{2}${]} & $13.40$       & $84.10$      & $4.21$         & $4.03$       & $110.67$  & $\bm{123.63}$      & $35.61$        & $31.91$   & $27.32$     & $9.96$     \\
		\textbf{Power} {[}mW{]} & $306.8$  & $\bm{23.77}$  &  $232$  &  $120.47$  &  $49.88$  &  $1075.2$  &  $123.55$  & $33.38$  &  $173.34$  &  $126.59$ \\
		\textbf{Energy} {[}pJ/bit{]}    & $12.56$       &   $2.16$           &   $27.95$             &    $29.45$          &     $\bm{2.10}$      &  $13.94$            &    $3.26$           &  $6.11$         &    $12.62$         &     $12.80$       \\				
		\bottomrule
	\end{tabular}
	\begin{tablenotes}
		\item[*] $^{\dagger}$ Scaled to $28$nm and $1.0$~V with area $\varpropto$ $s^2$, frequency $\varpropto$ $1/s$, and power $\varpropto$ $s\!\cdot\!u^{2}$, where $s$ is the scaling factor to $28$nm and $u$ is the scaling factor to $1.0$~V.
		\item[*] $^{\ddagger}$ We let the number of iterations be fixed to focus on the worst-case performance. 
		\item[*] $^{\diamond}$ For the missing voltages in~\cite{cui2020design,lin202133Gbps,yun2021area}, we assume these works all operate at $1.0$~V.
		\item[*] $^{*}$ In our 5G~NR LDPC decoder, the peak throughput is attained with a code configuration of BG1, $R=\frac{8}{9}$, and $Z=384$.
	\end{tablenotes}
\end{table*}

Table~\ref{tab:5GSynthesisPostlayout} provides both the synthesis results and post-layout results of our 5G~NR LDPC decoder.
The synthesis results indicate that our decoder has a cell area of $1.274$~mm$^{2}$ with a frequency of $1250$~MHz.
When all physical design processes (e.g., placement and routing) are done, the post-layout of our decoder has a core area of $1.823$~mm$^{2}$ with a maximum operating frequency of $895$~MHz.
For LDPC codes at BG1 with $R=\frac{8}{9}$ and $Z=384$, the implemented 5G~NR LDPC decoder (setting a fixed iteration value of $4$) achieves a peak throughput $\Theta_{\mathrm{T/P}}$ of~$24.42$~Gbps~as~follows:
\begin{equation}\label{eq:PeakTP}
	\begin{aligned}
		\Theta_{\mathrm{T/P}}&=\frac{Z_{\max}\cdot N_{p}}{\mathcal{L}}\cdot F\\
		&=\frac{384\times27}{380}\times0.895~\text{Gbps}=24.42~\text{Gbps},\\
	\end{aligned}
\end{equation}
where $\mathcal{L}\!=\!4\times\left(79\!+\!\left(19\!-\!3\right)\right)\!=\!380$~cycles determined by~\eqref{eq:latency5GLDPCadvanced} and $F$ is the operating frequency.
It is noteworthy that using the block-parallel architecture, our throughput~$\Theta_{\mathrm{T/P}}$ already satisfies the peak throughput requirement of $20$~Gbps as stipulated in the 5G standard~\cite{Hui2018VTMag}.

Fig.~\ref{fig:SecV_postlayout} illustrates the post-layout of our 5G~NR LDPC decoder, where the core size is $1.35\times 1.35$~mm$^{2}$ with a cell utilization of $72.9\%$.
When running in the case of~$\Theta_{\mathrm{T/P}}$, this decoder demonstrates a dynamic power of $306.8$~mW and an energy consumption of~$12.56$~pJ/bit.
Table~\ref{tab:PowerBreakdown} presents a comprehensive power breakdown of our decoder.
Notably, the power usage of the NCU logic only accounts for~$10.6\%$ of the total.
In addition, Fig.~\ref{fig:SecV_Plotresults} provides a detailed area analysis of cells.
Memory macros account for $60\%$ of the total cell area in our 5G~NR LDPC decoder, with the Q-memory macros, T-memory macros, and R-memory macros contributing around $14\%$, $14\%$, and $32\%$, respectively.
This substantial memory overhead in the 5G~NR LDPC decoder mitigates the impact of complex decoding algorithms on overall hardware efficiency.
Hence, within the context of achieving the peak throughput of $20$~Gbps, using more complex decoding algorithms is justified to further enhance the error-correcting performance of 5G~NR~LDPC~decoders.

As our decoder is compatible with all 5G~NR LDPC codes, the corresponding throughput and area efficiency vary depending on the code configurations.
With $Z_{\max}=384$ and the maximum $K_{u}$, we sweep all code rates of BG1 $\left(\frac{1}{3}\leq R\leq\frac{8}{9}\right)$ and BG2 $\left(\frac{1}{5}\leq R\leq\frac{2}{3}\right)$ at a fixed iteration value of $4$ and plot the corresponding throughput and area efficiency in Fig.~\ref{fig:SecV_Plotresults}.
For BG1 with $R=\frac{1}{3}$, our decoder achieves a throughput of $17.60$~Gbps and an area efficiency of $9.66$~Gbps/mm$^{2}$.
When the code rate increases to $\frac{8}{9}$, our decoder reaches a peak throughput of $24.42$~Gbps and a maximum area efficiency of $13.40$~Gbps/mm$^{2}$.
BG2 exhibits a similar trend to BG1, with corresponding peak throughput and area efficiency values of $24.34$~Gbps and $13.36$~Gbps/mm$^{2}$, respectively.
Moreover, our decoder can employ the PPCs as an early termination criterion to further enhance the average throughput.
For LDPC codes at BG1 with $R=\frac{8}{9}$ and $Z=384$, the average iteration and corresponding average throughput, under BPSK and $I_{\max}=15$, are~summarized~in~Table~\ref{tab:AvgTP}.

\subsection{Comparison With Previous Works}
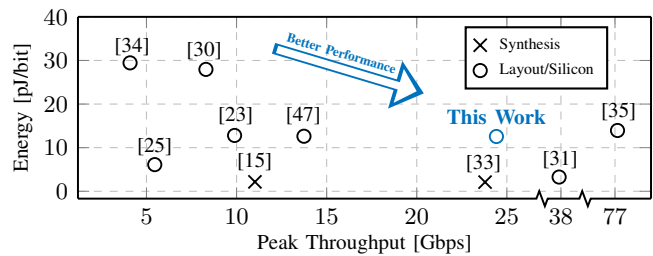
\begin{figure}[t]
	\centering
\pgfplotsset{compat=1.16}
\usetikzlibrary{arrows.meta, bending}

\begin{tikzpicture}
	\usepgflibrary{decorations.pathmorphing}
	\usepgflibrary[decorations.pathmorphing]
	\usetikzlibrary{decorations.pathmorphing}
	\usetikzlibrary[decorations.pathmorphing]
	
	\usetikzlibrary{arrows}
	\usetikzlibrary{shapes}
	
	\definecolor{myskyblue}{RGB}{39,156,191}
	\definecolor{myblued}{RGB}{0,114,189}
	\definecolor{myred}{RGB}{217,83,25}
	\definecolor{myredd}{RGB}{248,74,173}
	\definecolor{myyellow}{RGB}{237,137,32}
	\definecolor{mypurple}{RGB}{126,47,142}
	\definecolor{myblues}{RGB}{77,190,238}
	\definecolor{mygreen}{RGB}{21, 165, 112}
	\pgfplotsset{
		label style = {font=\fontsize{9pt}{7.2}\selectfont}, 
		tick label style = {font=\fontsize{9pt}{7.2}\selectfont} 
	}
	\begin{axis} [
		scale = 1,
		grid=both,
		height=4cm,
		width=9.2cm,
		ymax=40,
		xmax=33,
		xlabel = {Peak Throughput~[Gbps]},
		ylabel = {Energy~[pJ/bit]},
		ylabel style={align=center},
		xtick={5, 10, 15, 20, 25, 28, 31},
		xticklabels={$5$, $10$, $15$, $20$, $25$, $38$, $77$},
		ylabel style={yshift=-0.1cm, font=\footnotesize},
		xlabel style={yshift=+0.1cm, font=\footnotesize}, 
		ymajorgrids=true,
		xmajorgrids=true,
		grid style=dashed,
		separate axis lines,
		x axis line style= { draw opacity=0 },
		y axis line style= { draw opacity=0 },
		thick,
		axis line style={line width=0.75pt},
		legend style={
			anchor={center},
			cells={anchor=west},
			column sep= 1mm, 
			font=\fontsize{6pt}{6}\selectfont, 
			thick,     
			draw=black,
		},
		legend columns=1, 
		legend style={at={(0.8,0.6)}, anchor=south},
		]
		
		\addplot[
			color=black,
			only marks,
			very thick,
			mark = x,
			mark size=3.5pt,
			line width=0.25mm,
		]
		coordinates {
			(11.02, 2.16)
		};
		\addlegendentry{\text{Synthesis}}	
		
		\addplot[
			color=black,
			only marks,
			very thick,
			mark = o,
			mark size=2.5pt,
			line width=0.25mm,
		]
		coordinates {
			(8.3, 27.95)
		};
		\addlegendentry{\text{Layout/Silicon}}
		
		\addplot[
			color=black,
			very thick,
			mark = o,
			mark size = 2.5pt,
			line width=0.25mm,
		]
		coordinates {
			(4.09, 29.45)
		};		
		
		\addplot[
			color=myblued,
			very thick,
			mark = o,
			mark size=2.5pt,
			line width=0.25mm,
		]
		coordinates {
			(24.42, 12.56)
		};
		
		\addplot[
			color=black,
			very thick,
			mark = x,
			mark size=3.5pt,
			line width=0.25mm,
		]
		coordinates {
			(23.79, 2.10)
		};		
		
		\addplot[
			color=black,
			very thick,
			mark = o,
			mark size=2.5pt,
			line width=0.25mm,
		]
		coordinates {
			(31.14, 13.94)
		};		
		
		\addplot[
			color=black,
			very thick,
			mark = o,
			mark size=2.5pt,
			line width=0.25mm,
		]
		coordinates {
			(27.9, 3.26)
		};		
		
		\addplot[
			color=black,
			very thick,
			mark = o,
			mark size=2.5pt,
			line width=0.25mm,
		]
		coordinates {
			(5.46, 6.11)
		};		
		
		\addplot[
			color=black,
			very thick,
			mark = o,
			mark size=2.5pt,
			line width=0.25mm,
		]
		coordinates {
			(13.74, 12.62)
		};		
		
		\addplot[
			color=black,
			very thick,
			mark = o,
			mark size=2.5pt,
			line width=0.25mm,
		]
		coordinates {
			(9.89, 12.80)
		};
		
		\node[myblued, above] at (axis cs:24.42, 13){\footnotesize{\textbf{This Work}}};
		\node[above] at (axis cs:10.8, 2.5){\footnotesize{\cite{cui2020design}}};
		\node[above] at (axis cs:8, 28){\footnotesize{\cite{lin202133Gbps}}};
		\node[above] at (axis cs:4, 29){\footnotesize{\cite{Su2022SSCL}}};
		\node[above] at (axis cs:23.5, 2){\footnotesize{\cite{yun2021area}}};
		\node[above] at (axis cs:31, 14){\footnotesize{\cite{verma2022low}}};
		\node[above] at (axis cs:27.75, 3){\footnotesize{\cite{Lee2022TCASI}}};
		\node[above] at (axis cs:5.25, 6){\footnotesize{\cite{Roth2010ASSCC}}};
		\node[above] at (axis cs:13.5, 13){\footnotesize{\cite{kumawat2015high}}};
		\node[above] at (axis cs:9.75, 13){\footnotesize{\cite{zhang2010JSSC}}};
		
		\node[myblued, above, rotate=-16.5] at (axis cs:15.6, 28.5){\tiny{\textbf{Better Performance}}};

		\path[-] (rel axis cs:0,0)     coordinate(botstart)
		--(rel axis cs:0.8,0)coordinate(interruptbotA)
		(rel axis cs:0.825,0)  coordinate(interruptbotB)
		--(rel axis cs:0.875,0)coordinate(interruptbotC)
		(rel axis cs:0.90,0)  coordinate(interruptbotD)
		--(rel axis cs:1,0)   coordinate(botstop);
		
		\path[-] (rel axis cs:0,1)     coordinate(topstart)
		--(rel axis cs:1,1)   coordinate(topstop);

		\path[-] (rel axis cs:0,0)     coordinate(leftstart)
		--(rel axis cs:0,1)   coordinate(leftstop);

		\path[-] (rel axis cs:1,0)     coordinate(rightstart)
		--(rel axis cs:1,1)   coordinate(rightstop);

	\end{axis}
	
	\draw[line width=0.75pt](botstart)-- (interruptbotA) decorate[decoration={zigzag,segment length = 1.5mm, amplitude = 1mm}]{--(interruptbotB)} -- (interruptbotC) decorate[decoration={zigzag,segment length = 1.5mm, amplitude = 1mm}]{--(interruptbotD)} -- (botstop);
	\draw[line width=0.75pt](topstart)-- (topstop);
	\draw[line width=0.75pt](leftstart)-- (leftstop);
	\draw[line width=0.75pt](rightstart)-- (rightstop);
	
	\path[draw=myblued,   line width=1.5mm, -{Triangle[length=4mm]}]     (2.6, 2.0) to    (4.6, 1.4);
	\path[draw=white, line width=1mm, -{Triangle[length=2.5mm]}, shorten >=1mm, shorten <=0.5mm]    	(2.6, 2.0) to   (4.6, 1.4);
	
\end{tikzpicture}
	\caption{Peak throughput vs. energy (normalized to $28$nm) of various SOA LDPC decoders.}
	\label{fig:hw_comp}
\end{figure}

Table~\ref{tab:5Gresults} provides a detailed comparison between our 5G~NR LDPC decoder with the SOA decoder implementations in~\cite{cui2020design,lin202133Gbps,yun2021area,Su2022SSCL,verma2022low,Lee2022TCASI,Roth2010ASSCC,kumawat2015high,zhang2010JSSC}.
To ensure fairness, we normalize all previous works to a $28$nm process with a supply voltage of $1.0$~V and set a fixed number of iterations to $4$.
Note that there is no early termination in Table~\ref{tab:5Gresults} to focus on the architecture.
Compared to a similar block-parallel 5G~NR LDPC decoder in~\cite{lin202133Gbps}, our work has a $2.94\times$ peak throughput, a $3.18\times$ area efficiency, and $55\%$ less energy consumption.
When compared to the SOA row-parallel architectures presented in~\cite{cui2020design,Su2022SSCL,yun2021area,verma2022low}, our decoder achieves a throughput that is $2.22\times$ faster than~\cite{cui2020design} and $1.03\times$ faster than~\cite{yun2021area}. 
Moreover, it demonstrates $3.32\times$ greater area efficiency than~\cite{Su2022SSCL} and consumes $9.9\%$ less energy than~\cite{verma2022low}.
Although the area overhead of these row-parallel 5G~NR LDPC decoders~\cite{cui2020design,Su2022SSCL,yun2021area,verma2022low} is better than our results, their maximum code lengths are much shorter than $N=26112$ required by the 5G standard, granting them a significant area advantage.
Indeed, these row-parallel architectures will suffer from high routing complexity to be compatible with all 5G~NR LDPC codes.
In comparison with the 5G~NR LDPC decoder in~\cite{Lee2022TCASI}, our peak throughput is $35.6\%$ inferior to~\cite{Lee2022TCASI}, but the PRP architecture of~\cite{Lee2022TCASI} has long decoding latency at medium to low code rates.
For instance, for LDPC codes at BG1 with $R=\frac{1}{3}$ and $Z=384$, our decoder can yield $17.60$~Gbps at a fixed iteration value of $4$, but the normalized throughput of~\cite{Lee2022TCASI} is only $9.74$~Gbps (calculated by (5) in~\cite{Lee2022TCASI}).
It is noteworthy that our GA-MS-3 decoder has a lower error-rate than the OMS decoder in~\cite{Lee2022TCASI}, as shown in Fig.~\ref{fig:MQAM}.
Given that the area overhead (e.g., routing and storage complexity) of LDPC decoders is largely influenced by varying maximum code lengths that can be processed and therefore difficult to compare, we only plot the peak throughput against energy in Fig.~\ref{fig:hw_comp} for various SOA LDPC decoders listed in Table~\ref{tab:5Gresults}.
From Fig.~\ref{fig:hw_comp}, our decoder outperforms many existing decoders but is still inferior to~\cite{Lee2022TCASI}.
Nonetheless, as mentioned before, our decoder achieves higher throughput than~\cite{Lee2022TCASI} at medium to low code rates and maintains a lower error-rate.
In conclusion, our decoder has an energy of $12.56$~pJ/bit, consuming $55\%$, $57.4\%$, and $9.9\%$ less than~\cite{lin202133Gbps,Su2022SSCL,verma2022low}, respectively.
This work also achieves a peak throughput of $24.42$~Gbps which is $2.22\times$, $2.94\times$, $5.97\times$, $1.03\times$, $4.47\times$, $1.78\times$, and $2.47\times$ faster than the SOA LDPC decoders \cite{cui2020design}, \cite{lin202133Gbps}, \cite{Su2022SSCL}, \cite{yun2021area}, \cite{Roth2010ASSCC}, \cite{kumawat2015high}, \cite{zhang2010JSSC}.
Moreover, the maximum area efficiency in our 5G~NR decoder is $13.40$~Gbps/mm$^{2}$, which is $3.18\times$, $3.32\times$, and $1.35\times$ higher than \cite{lin202133Gbps}, \cite{Su2022SSCL}, \cite{zhang2010JSSC}, respectively.

\section{Conclusions}\label{sec:SecVI_conclusion}
In this paper, we propose high-performance and low-complexity GA-MS decoding.
By truncating the number of incoming messages in CN processing, we can make a trade-off between error-correcting performance and computational complexity.
By incorporating the well-designed LUTs, quantization schemes, and other approximation techniques, our fixed-point GA-MS decoding exhibits only a minor gap of $0.1$~dB compared to floating-point SP decoding under various 5G~NR code configurations and high-order modulations.
We also present a fully reconfigurable 5G~NR LDPC decoder implementation, compatible with all 5G~NR LDPC codes.
The $28$nm FD-SOI post-layout implementation results show that our decoder has a core area of $1.823$~mm$^{2}$, achieves a peak throughput of $24.42$~Gbps at $895$~MHz, and has an energy consumption of $12.56$~pJ/bit with a supply voltage of $1.0$~V.

\bibliographystyle{IEEEtran}
\bibliography{IEEEabrv, bibliography}

\begin{thebibliography}{10}
\providecommand{\url}[1]{#1}
\csname url@samestyle\endcsname
\providecommand{\newblock}{\relax}
\providecommand{\bibinfo}[2]{#2}
\providecommand{\BIBentrySTDinterwordspacing}{\spaceskip=0pt\relax}
\providecommand{\BIBentryALTinterwordstretchfactor}{4}
\providecommand{\BIBentryALTinterwordspacing}{\spaceskip=\fontdimen2\font plus
\BIBentryALTinterwordstretchfactor\fontdimen3\font minus
  \fontdimen4\font\relax}
\providecommand{\BIBforeignlanguage}[2]{{%
\expandafter\ifx\csname l@#1\endcsname\relax
\typeout{** WARNING: IEEEtran.bst: No hyphenation pattern has been}%
\typeout{** loaded for the language `#1'. Using the pattern for}%
\typeout{** the default language instead.}%
\else
\language=\csname l@#1\endcsname
\fi
#2}}
\providecommand{\BIBdecl}{\relax}
\BIBdecl

\bibitem{gallager1962low}
R.~Gallager, ``Low-density parity-check codes,'' \emph{IRE Trans. Inf. Theory},
  vol.~8, no.~1, pp. 21--28, Jan. 1962.

\bibitem{ATSC2007}
\emph{Standard: Synchronization standard for distributed transmission},
  {Advanced Television System Committee (ATSC)}, Feb. 2007.

\bibitem{80211n2008LDPC}
\emph{Wireless LAN medium access control {(MAC)} and physical layer {(PHY)}
  specifications: Enhancements for higher throughput}, {IEEE P802.11n/D5.02,
  Part 11}, Jul. 2008.

\bibitem{DVB2009}
\emph{Digital video broadcasting {(DVB)} user guidelines for the second
  generation system for broadcasting, interactive services, news gathering and
  other broadband satellite applications {(DVB-S2)}}, {ETSI TR 102 376}, Feb.
  2009.

\bibitem{5Gstandard2016}
\emph{{Chairman’s notes of AI 7.1.5 on consideration on LDPC design for NR}},
  {3GPP R1-1611112 Release 16}, Nov. 2016.

\bibitem{5Gstandard2018}
\emph{{5G NR}: multiplexing and channel coding}, 3GPP TS 38.212 version 15.2.0
  Release 15, Jul. 2018.

\bibitem{FR2001FGTIT}
F.~Kschischang, B.~Frey, and H.-A. Loeliger, ``Factor graphs and the
  sum-product algorithm,'' \emph{{IEEE} Trans. Inf. Theory}, vol.~47, no.~2,
  pp. 498--519, Feb. 2001.

\bibitem{Urbanke2001TIT}
T.~Richardson and R.~Urbanke, ``The capacity of low-density parity-check codes
  under message-passing decoding,'' \emph{{IEEE} Trans. Inf. Theory}, vol.~47,
  no.~2, pp. 599--618, Feb. 2001.

\bibitem{Mansour2003HTVLSI}
M.~Mansour and N.~Shanbhag, ``High-throughput {LDPC} decoders,'' \emph{{IEEE}
  Trans. {VLSI} Syst.}, vol.~11, no.~6, pp. 976--996, Dec. 2003.

\bibitem{wiberg1996codes}
N.~Wiberg, ``Codes and decoding on general graphs,'' 1996.

\bibitem{fossorier1999reduced}
M.~P. Fossorier, M.~Mihaljevic, and H.~Imai, ``Reduced complexity iterative
  decoding of low-density parity check codes based on belief propagation,''
  \emph{{IEEE} Trans. Commun.}, vol.~47, no.~5, pp. 673--680, May 1999.

\bibitem{chen2005TCOM}
J.~Chen, A.~Dholakia, E.~Eleftheriou, M.~Fossorier, and X.-Y. Hu,
  ``Reduced-complexity decoding of {LDPC} codes,'' \emph{{IEEE} Trans.
  Commun.}, vol.~53, no.~8, pp. 1288--1299, Aug. 2005.

\bibitem{wu2010adaptive}
X.~Wu, Y.~Song, M.~Jiang, and C.~Zhao, ``Adaptive-normalized/offset min-sum
  algorithm,'' \emph{{IEEE} Commun. Lett.}, vol.~14, no.~7, pp. 667--669, Jul.
  2010.

\bibitem{le2019adaptation}
K.~Le~Trung, F.~Ghaffari, and D.~Declercq, ``An adaptation of min-sum decoder
  for {5G} low-density parity-check codes,'' in \emph{Proc. IEEE Int. Symp.
  Circuits Syst.}, 2019, pp. 1--5.

\bibitem{cui2020design}
H.~Cui, F.~Ghaffari, K.~Le, D.~Declercq, J.~Lin, and Z.~Wang, ``Design of
  high-performance and area-efficient decoder for {5G LDPC} codes,''
  \emph{{IEEE} Trans. Circuits Syst. {I}}, vol.~68, no.~2, pp. 879--891, Feb.
  2020.

\bibitem{savin2008self}
V.~Savin, ``Self-corrected min-sum decoding of {LDPC} codes,'' in \emph{Proc.
  IEEE Int. Symp. Inf. Theory}, 2008, pp. 146--150.

\bibitem{zhang2006two}
J.~Zhang, M.~Fossorier, and D.~Gu, ``Two-dimensional correction for min-sum
  decoding of irregular {LDPC} codes,'' \emph{{IEEE} Commun. Lett.}, vol.~10,
  no.~3, pp. 180--182, Mar. 2006.

\bibitem{kang2020enhanced}
P.~Kang, Y.~Xie, L.~Yang, and J.~Yuan, ``Enhanced quasi-maximum likelihood
  decoding based on {2D} modified min-sum algorithm for {5G} {LDPC} codes,''
  \emph{{IEEE} Trans. Commun.}, vol.~68, no.~11, pp. 6669--6682, Nov. 2020.

\bibitem{jones2003approximate}
C.~Jones, E.~Valles, M.~Smith, and J.~Villasenor, ``Approximate-min constraint
  node updating for {LDPC} code decoding,'' in \emph{Proc. IEEE Military Comm.
  Conf.}, vol.~1, 2003, pp. 157--162.

\bibitem{zhou2019generalized}
W.~Zhou and M.~Lentmaier, ``Generalized two-magnitude check node updating with
  self correction for {5G LDPC} codes decoding,'' in \emph{Proc. IEEE Int.
  Conf. on Syst. Comm. Coding}, 2019, pp. 1--6.

\bibitem{Adj2017OPatent}
T.~J. Richardson, S.~Kudekar, and V.~Loncke, \emph{Adjusted mim-sum decoder},
  US Patent, Apr. 2017.

\bibitem{kuo2008flexible}
T.-C. Kuo and A.~N. Willson, ``A flexible decoder {IC} for {WiMAX} {QC-LDPC}
  codes,'' in \emph{Proc. IEEE Custom Integrated Circuits Conf.}, 2008, pp.
  527--530.

\bibitem{zhang2010JSSC}
Z.~Zhang, V.~Anantharam, M.~J. Wainwright, and B.~Nikolic, ``An efficient
  {10GBASE-T} {Ethernet} {LDPC} decoder design with low error floors,''
  \emph{{IEEE} J. Solid-State Circuits}, vol.~45, no.~4, pp. 843--855, Apr.
  2010.

\bibitem{Studer2008Asilomar}
C.~Studer, N.~Preyss, C.~Roth, and A.~Burg, ``Configurable high-throughput
  decoder architecture for quasi-cyclic {LDPC} codes,'' in \emph{Proc. IEEE
  Asilomar Conf. Signals, Syst. Compt.}, 2008, pp. 1137--1142.

\bibitem{Roth2010ASSCC}
C.~Roth, P.~Meinerzhagen, C.~Studer, and A.~Burg, ``A 15.8 {pJ/bit/iter}
  quasi-cyclic {LDPC} decoder for {IEEE} 802.11n in {90~nm} {CMOS},'' in
  \emph{Proc. IEEE Asian Solid-State Circuits Conf.}, 2010, pp. 1--4.

\bibitem{Hui2018VTMag}
D.~Hui, S.~Sandberg, Y.~Blankenship, M.~Andersson, and L.~Grosjean, ``Channel
  coding in {5G} new radio: A tutorial overview and performance comparison with
  {4G LTE},'' \emph{{IEEE} Veh. Technol. Mag.}, vol.~13, no.~4, pp. 60--69,
  Dec. 2018.

\bibitem{blanksby2002690}
A.~J. Blanksby and C.~J. Howland, ``A {690-mW} 1-{Gb/s} 1024-b, rate-1/2
  low-density parity-check code decoder,'' \emph{{IEEE} J. Solid-State
  Circuits}, vol.~37, no.~3, pp. 404--412, Mar. 2002.

\bibitem{cheng2014fully}
C.-C. Cheng, J.-D. Yang, H.-C. Lee, C.-H. Yang, and Y.-L. Ueng, ``A fully
  parallel {LDPC} decoder architecture using probabilistic min-sum algorithm
  for high-throughput applications,'' \emph{{IEEE} Trans. {VLSI} Syst.},
  vol.~61, no.~9, pp. 2738--2746, Sep. 2014.

\bibitem{ghanaatian2017588}
R.~Ghanaatian, A.~Balatsoukas-Stimming, T.~C. M{\"u}ller, M.~Meidlinger,
  G.~Matz, A.~Teman, and A.~Burg, ``A 588-{Gb/s} {LDPC} decoder based on
  finite-alphabet message passing,'' \emph{{IEEE} Trans. {VLSI} Syst.},
  vol.~26, no.~2, pp. 329--340, Feb. 2018.

\bibitem{lin202133Gbps}
C.-Y. Lin, L.-W. Liu, Y.-C. Liao, and H.-C. Chang, ``A 33.2 {Gbps/iter.}
  reconfigurable {LDPC} decoder fully compliant with {5G NR} applications,'' in
  \emph{Proc. IEEE Int. Symp. Circuits Syst.}, 2021, pp. 1--5.

\bibitem{Lee2022TCASI}
S.~Lee, S.~Park, B.~Jang, and I.-C. Park, ``Multi-mode {QC-LDPC} decoding
  architecture with novel memory access scheduling for {5G} {New-Radio}
  standard,'' \emph{{IEEE} Trans. Circuits Syst. {I}}, vol.~69, no.~5, pp.
  2035--2048, May 2022.

\bibitem{nadal2021parallel}
J.~Nadal and A.~Baghdadi, ``Parallel and flexible {5G} {LDPC} decoder
  architecture targeting {FPGA},'' \emph{{IEEE} Trans. {VLSI} Syst.}, vol.~29,
  no.~6, pp. 1141--1151, Jun. 2021.

\bibitem{yun2021area}
S.~Yun, B.~Y. Kong, and Y.~Lee, ``Area-and energy-efficient {LDPC} decoder
  using mixed-resolution check-node processing,'' \emph{{IEEE} Trans. Circuits
  Syst. {II}}, vol.~69, no.~3, pp. 999--1003, Mar. 2022.

\bibitem{Su2022SSCL}
B.-S. Su, C.-H. Lee, and T.-D. Chiueh, ``A 58.6/91.3 {pJ/b} dual-mode
  belief-propagation decoder for {LDPC} and polar codes in the {5G}
  communications standard,'' \emph{IEEE Solid-State Circuits Lett.}, vol.~5,
  Apr. 2022.

\bibitem{verma2022low}
A.~Verma and R.~Shrestha, ``Low computational-complexity {SOMS}-algorithm and
  high-throughput decoder architecture for {QC-LDPC} codes,'' \emph{{IEEE}
  Trans. Veh. Technol.}, vol.~72, no.~1, pp. 66--80, Jan. 2023.

\bibitem{tanner1981recursive}
R.~Tanner, ``A recursive approach to low complexity codes,'' \emph{{IEEE}
  Trans. Inf. Theory}, vol.~27, no.~5, pp. 533--547, Sep. 1981.

\bibitem{fossorier2004quasicyclic}
M.~P. Fossorier, ``Quasi-cyclic low-density parity-check codes from circulant
  permutation matrices,'' \emph{{IEEE} Trans. Inf. Theory}, vol.~50, no.~8, pp.
  1788--1793, Aug. 2004.

\bibitem{zhong2005block}
H.~Zhong and T.~Zhang, ``Block-{LDPC}: A practical {LDPC} coding system design
  approach,'' \emph{{IEEE} Trans. Circuits Syst. {I}}, vol.~52, no.~4, pp.
  766--775, Apr. 2005.

\bibitem{Zhon2020TCASII}
Z.~Zhong, Y.~Huang, Z.~Zhang, X.~You, and C.~Zhang, ``A flexible and high
  parallel permutation network for {5G} {LDPC} decoders,'' \emph{{IEEE} Trans.
  Circuits Syst. {II}}, vol.~67, no.~12, pp. 3018--3022, Jun. 2020.

\bibitem{sharon2004efficient}
E.~Sharon, S.~Litsyn, and J.~Goldberger, ``An efficient message-passing
  schedule for {LDPC} decoding,'' in \emph{Proc. IEEE Conven. Elect. Electron.
  Eng. Isreal}, 2004, pp. 223--226.

\bibitem{hocevar2004reduced}
D.~E. Hocevar, ``A reduced complexity decoder architecture via layered decoding
  of {LDPC} codes,'' in \emph{Proc. IEEE Workshop Signal Process. Syst.}, 2004,
  pp. 107--112.

\bibitem{mao2016low}
J.~Mao, M.~A. Abdullahi, P.~Xiao, and A.~Cao, ``A low complexity {256QAM} soft
  demapper for {5G} mobile system,'' in \emph{Proc. IEEE Euro. Conf. Netw.
  Commun.}, 2016, pp. 16--21.

\bibitem{le2020low}
B.~Le~Gal, Y.~Delomier, C.~Leroux, and C.~J{\'e}go, ``Low-latency sorter
  architecture for polar codes successive-cancellation-list decoding,'' in
  \emph{Proc. IEEE Workshop Signal Process. Syst.}, 2020, pp. 1--5.

\bibitem{Chang2021Dynamic}
T.~C.-Y. Chang, P.-H. Wang, J.-J. Weng, I.-H. Lee, and Y.~T. Su,
  ``Belief-propagation decoding of {LDPC} codes with variable node–centric
  dynamic schedules,'' \emph{{IEEE} Trans. Commun.}, vol.~69, no.~8, pp.
  5014--5027, Aug. 2021.

\bibitem{liang2019hardware}
C.-Y. Liang, M.-R. Li, H.-C. Lee, H.-Y. Lee, and Y.-L. Ueng,
  ``Hardware-friendly {LDPC} decoding scheduling for {5G} {HARQ}
  applications,'' in \emph{Proc. IEEE Int. Conf. Acoust. Speech Signal
  Process.}, 2019, pp. 1418--1422.

\bibitem{tian2022novel}
K.~Tian and H.~Wang, ``A novel base graph based static scheduling scheme for
  layered decoding of {5G} {LDPC} codes,'' \emph{{IEEE} Commun. Lett.},
  vol.~26, no.~7, pp. 1450--1453, Jul. 2022.

\bibitem{kumawat2015high}
S.~Kumawat, R.~Shrestha, N.~Daga, and R.~Paily, ``High-throughput
  {LDPC}-decoder architecture using efficient comparison techniques \& dynamic
  multi-frame processing schedule,'' \emph{{IEEE} Trans. Circuits Syst. {I}},
  vol.~62, no.~5, pp. 1421--1430, May 2015.

\end{thebibliography}

\end{document}